\documentclass[11pt]{article}

\usepackage[T1]{fontenc}
\usepackage{amsmath}
\usepackage{enumitem}
\usepackage{amssymb}
\usepackage{amsthm}
\usepackage{aliascnt}
\usepackage{mathtools}
\usepackage{microtype}
\usepackage{etoolbox}
\usepackage[margin=1in]{geometry}
\usepackage[bottom]{footmisc}
\usepackage{xcolor}
\usepackage[pagebackref=true]{hyperref}
\usepackage[capitalise]{cleveref}
\usepackage{verbatim}
\usepackage{titlesec}
\usepackage{todonotes}

\newif\ifdraft
\draftfalse
\ifdraft
\newcommand{\omar}[1]{{
    \footnotesize
    [\textbf{\textcolor{orange}{#1}}
    \textcolor{orange!70!black}{--omar}]
}}
\newcommand{\rohan}[1]{{
    \footnotesize
    [\textbf{\textcolor{red}{#1}}
    \textcolor{red!70!black}{--rohan}]
}}
\newcommand{\venkat}[1]{{
    \footnotesize
    [\textbf{\textcolor{blue}{#1}}
    \textcolor{blue!60!black}{--venkat}]
}}
\else
\newcommand{\omar}[1]{}
\newcommand{\rohan}[1]{}
\newcommand{\venkat}[1]{}
\fi
\titlespacing*{\paragraph}
    {0pt}{1ex}{1em}

\renewcommand{\le}{\leqslant}
\renewcommand{\leq}{\leqslant}
\renewcommand{\ge}{\geqslant}
\renewcommand{\geq}{\geqslant}

\allowdisplaybreaks

\hypersetup{
    colorlinks=true,
    linkcolor=purple!70!black,
    citecolor=purple!70!black,
    urlcolor=purple!70!black,
    pdftitle={Not every Reed--Solomon code is list-decodable beyond the Johnson radius}
}

\theoremstyle{plain}

\newtheorem{theorem}{Theorem}[section]
\newtheorem*{theorem*}{Theorem}

\newaliascnt{prop}{theorem}
\newtheorem{prop}[prop]{Proposition}
\aliascntresetthe{prop}

\newaliascnt{fact}{theorem}

\aliascntresetthe{fact}

\newaliascnt{lemma}{theorem}
\newtheorem{lemma}[lemma]{Lemma}
\aliascntresetthe{lemma}

\newaliascnt{corollary}{theorem}
\newtheorem{corollary}[corollary]{Corollary}
\aliascntresetthe{corollary}

\theoremstyle{definition}
\newaliascnt{definition}{theorem}
\newtheorem{definition}[definition]{Definition}
\aliascntresetthe{definition}

\theoremstyle{remark}
\newaliascnt{remark}{theorem}

\aliascntresetthe{remark}

\usepackage{thm-restate}

\AtEndEnvironment{remark}{\hfill$\triangleleft$}

\newtheoremstyle{claimstyle}
  {\topsep}{\topsep}{\normalfont}{0pt}{\bfseries}{.}{.5em}
  {\thmname{#1}\thmnumber{ #2}\thmnote{ (#3)}}
\theoremstyle{claimstyle}
\newtheorem{claim}{Claim}[subsection]

\newenvironment{claimproof}
  {\begin{proof}[Proof of claim]}
  {\end{proof}}

\crefname{theorem}{Theorem}{Theorems}
\crefname{prop}{Proposition}{Propositions}
\crefname{fact}{Fact}{Facts}
\crefname{lemma}{Lemma}{Lemmas}
\crefname{corollary}{Corollary}{Corollaries}
\crefname{definition}{Definition}{Definitions}
\crefname{remark}{Remark}{Remarks}
\crefname{claim}{Claim}{Claims}

\newcommand{\F}{\mathbb{F}}
\newcommand{\Tr}{\operatorname{Tr}}
\newcommand{\RS}{\operatorname{RS}}
\newcommand{\dist}{d_{\mathrm H}}
\newcommand{\eps}{\varepsilon}
\newcommand{\abs}[1]{\left|#1\right|}
\newcommand{\ceil}[1]{\left\lceil #1 \right\rceil}
\newcommand{\Ball}[2]{B\!\left(#1,#2\right)}

\title{Combinatorial limitations of list-decoding Reed--Solomon codes}
\author{Omar Alrabiah\thanks{Department of Electrical Engineering and Computer Sciences, UC Berkeley, Berkeley, CA 94709, USA. Email: \href{mailto:oalrabiah@berkeley.edu}{\texttt{oalrabiah@berkeley.edu}}. Research supported in part by a Saudi Arabian Cultural Mission (SACM) Scholarship, ONR grant N00014-24-1-2491, and V.~Guruswami's Simons Investigator Award.} \and Rohan Goyal\thanks{Massachusetts Institute of Technology, Cambridge, MA, USA. Email: \href{mailto:rohan_g@mit.edu}{\texttt{rohan\_g@mit.edu}}.  Research supported by Yael Tauman Kalai's NSF Award 6956054.} \and Venkatesan Guruswami\thanks{Department of Electrical Engineering and Computer Sciences, Department of Mathematics, and the Simons Institute for the Theory of Computing, UC Berkeley, Berkeley, CA 94709, USA. Email: \href{mailto:venkatg@berkeley.edu}{\texttt{venkatg@berkeley.edu}}. Research supported in part by a Simons Investigator Award, a UC Noyce initiative award, DOD Advanced
Research Projects Agency grant HR0011262E031, and NSF grant 2211972.}}
\date{October 2026}

\begin{document}
\maketitle
\thispagestyle{empty}

\begin{abstract}
For infinitely many rates $R \in (0,1)$, we show that full-length Reed--Solomon codes over fields of low characteristic can contain super-polynomially many codewords in a Hamming ball of radius just exceeding the Johnson radius. Over prime fields, we exhibit full-length Reed--Solomon codes of constant rate $R$ with polynomially many codewords in a Hamming ball of relative radius bounded strictly below $1-R$. This refutes a recent conjecture that, over prime fields, list sizes below capacity are always bounded by a constant. Together, these results demonstrate that certain restrictions in recent works on explicit Reed--Solomon codes achieving list-decoding capacity, related to the choice of field or evaluation points or to the list-size guarantee, are inherent. 
\end{abstract}

\newpage
\section{Introduction}
\label{sec:introduction}

List decoding, introduced by Elias and Wozencraft~\cite{Eli57,Woz58}, is a decoding paradigm in which a decoder for a code $C \subseteq \Sigma^n$ over a finite alphabet $\Sigma$ receives a word $y \in \Sigma^n$ and outputs all codewords $c \in C$ satisfying $\dist(c,y) \leq \rho n$, where $\dist$ denotes Hamming distance and $\rho$ is the prescribed fractional decoding radius. Unlike unique decoding, which seeks a single codeword close to the received word, list decoding remains meaningful beyond half the minimum distance of $C$, where a received word may be close to multiple codewords. More formally, we call $C$ \emph{$(\rho,L)$-list-decodable} if $\abs{C \cap \Ball{y}{\rho n}} \leq L$ for every $y \in \Sigma^n$, where $\Ball{y}{r} \coloneqq \{z \in \Sigma^n : \dist(z,y) \leq r\}$ denotes the Hamming ball of radius $r$ around $y$. A central goal in list-decoding is to maximize the decoding radius $\rho$ while keeping the list size $L$ small, ideally by a constant or at most a polynomial in $n$. For arbitrary codes, a first general answer to this question is provided by the Johnson bound~\cite{Joh62} (for simple proofs and extensions, see~\cite{GS-johnson}). Specifically, if a code $C$ has relative distance $\delta$, then the list size $\abs{C \cap \Ball{y}{\rho n}}$ is at most polynomial in $n$ for every $y \in \Sigma^n$ whenever $\rho < 1-\sqrt{1-\delta}$, and it is in fact $O(1/\eps)$ when $\rho = 1-\sqrt{1-\delta}-\eps$. We call $1-\sqrt{1-\delta}$ the \emph{Johnson radius}. 

A canonical family of codes in this context is Reed--Solomon codes~\cite{RS60}, whose optimal distance makes them especially well suited to the study of list-decoding. A Reed--Solomon code of dimension $k$ over a finite field $\F$ encodes each polynomial in $\F[X]$ of degree less than $k$ by its values at $n$ distinct evaluation points in $\F$. It has rate $R = k/n$ and relative distance $\delta = (n-k+1)/n = 1-R+1/n$, which meets the Singleton bound~\cite{Sin64}. Consequently, since $1-\delta = R-1/n$, its Johnson radius equals $1-\sqrt{R-1/n}$, which, as a function of the rate $R$, is $1-\sqrt{R}$ up to an additive $O(1/n)$ term. Thus, this optimal distance translates into a strong list-decoding guarantee, and this is made particularly compelling by the existence of \emph{efficient} algorithms that achieve it. This began with classical algorithms~\cite{Peterson60,Ber68,Massey69,WB86} that decode Reed--Solomon codewords in polynomial time up to half their minimum distance and was later extended to the list-decoding algorithm of Guruswami and Sudan~\cite{GS99}, building on Sudan's earlier algorithm~\cite{Sud97}, that goes up to the Johnson radius for \emph{arbitrary} Reed--Solomon codes.

In general, however, super-polynomial list sizes can occur just beyond the Johnson radius. A folklore random coding argument shows that there are (not necessarily linear) codes $C$ of relative distance $\delta$ and words $y$ for which $\abs{C \cap \Ball{y}{\rho n}}$ is super-polynomial in $n$, with $\rho$ only slightly exceeding $1-\sqrt{1-\delta}$~\cite[Section~4.3]{GRS}. Moreover, this phenomenon can occur even for binary \emph{linear} codes~\cite{Gur02,GS03}.\footnote{For binary codes, the Johnson radius is $(1-\sqrt{1-2\delta})/2$, and more generally, for alphabet size $q$, it is $(1-1/q)(1-\sqrt{1-q\delta/(q-1)})$.}

These limitations raise the question of how far beyond the Johnson radius Reed--Solomon codes can be efficiently list-decoded. The ultimate limit is the \emph{list-decoding capacity} $1-R$, beyond which no code of rate $R$ can have sub-exponential list sizes. Indeed, if $\abs{\Sigma} \geq n$, say, then for every $\eps > 0$ the product $\abs{C}\cdot\abs{\Ball{y}{(1-R+\eps)n}}$ exceeds $\abs{\Sigma}^n$ by an exponential factor, and hence $\abs{C \cap \Ball{y}{(1-R+\eps)n}}$ is exponential in $n$ on average over a uniformly random center $y \in \Sigma^n$. On the other hand, random codes of rate $R$ over large alphabets achieve the capacity, being $(1-R-\eps,O(1/\eps))$-list-decodable~\cite{ZP81,Eli91}.

For Reed--Solomon codes, however, it remained conceivable that \emph{every} such code could be list-decoded beyond the Johnson radius, and in fact up to the capacity $1-R$, with polynomially bounded lists. More precisely, one could hope that for every $\eps > 0$, every Reed--Solomon code of rate $R$ is $(1-R-\eps,L)$-list-decodable with $L$ polynomial in $n$. Some evidence for this possibility comes from Reed--Solomon codes with \emph{random} (or generic) evaluation points over large fields, which exhibit behavior similar to that of random linear codes.  Starting with the work of Rudra and Wootters~\cite{RW14} in the high-noise regime, and later that of Shangguan and Tamo~\cite{ST20}, a sequence of works established list-decodability of such codes beyond the Johnson radius~\cite{ST20,GLSTW21,FKS22,GST23}, eventually reaching capacity with constant list size~\cite{BGM23,GZ23,AGL24b,AGGLZ25}. However, these results apply to random evaluation points and provide no guarantees for a particular Reed--Solomon code, such as the \emph{full-length} code over a given field, whose evaluation points are all the elements of the field and whose length $n$ is therefore the field size. For arbitrary evaluation points, the recent work of Brakensiek, Chen, Putterman, Zhang and Zheng~\cite{BCPZZ26} establishes capacity-approaching list-decoding over sufficiently large prime fields, as we discuss below.

The contrasting possibility---that the Johnson radius is the true limit for some Reed--Solomon codes---has been explored in several earlier works. This began with the work of Justesen and H{\o}holdt~\cite{JH01}, which studied the list sizes of MDS codes (codes meeting the Singleton bound, such as Reed--Solomon codes) around the Johnson radius, proving upper bounds on the number of codewords and exhibiting received words with polynomially many nearby codewords. In a related direction, Guruswami and Rudra~\cite{GR06} considered the notion of list recovery, a generalization of list decoding in which each coordinate of the received word is replaced by a small set of allowed symbols, and showed that Reed--Solomon codes cannot be list-recovered beyond the natural analogue of the Johnson radius with polynomially bounded lists. Their construction uses polynomials over an extension field that take values in a subfield, an ingredient that also appears in our proof of \cref{thm:near-johnson-intro}. 

Complementing these combinatorial investigations, several works have established computational barriers to decoding Reed--Solomon codes. In particular, maximum-likelihood decoding is NP-hard in general~\cite{GV05}, and in certain parameter regimes, list-decoding beyond the Johnson radius is at least as hard as computing discrete logarithms in finite fields~\cite{CW07}. These hardness results, however, do not rule out polynomially bounded list sizes.

An earlier precursor to our work, more directly addressing the existence of large lists beyond the Johnson radius, is the work of Ben-Sasson, Kopparty and Radhakrishnan~\cite{BKR10}, who used subspace polynomials to construct received words with super-polynomially many nearby codewords in full-length Reed--Solomon codes beyond the Johnson radius. Their construction, however, yields super-polynomial list sizes only when the code rate tends to zero, a regime in which both the Johnson radius and the list-decoding capacity approach $1$. At any fixed positive rate, the list sizes obtained by their construction are only polynomial in the block length.

\paragraph{Extension field RS codes that are not list-decodable just past the Johnson radius.}
Our first result shows that the Johnson radius is essentially a barrier to list decoding with polynomially bounded lists for certain full-length Reed--Solomon codes of constant rate over extension fields. Specifically, for rates $R = q^{-2}$ and $R = (1-q^{-1})^2$, where $q$ is a prime power, we construct received words with super-polynomially many nearby codewords at a relative distance arbitrarily close to the Johnson radius $1-\sqrt{R}$ from above. In particular, there is a received word $y$ such that for every fixed $\eps > 0$, these codes have super-polynomially many codewords in the Hamming ball $\Ball{y}{(1-\sqrt{R}+\eps)n}$. Unlike the earlier constructions based on subspace polynomials, our lower bounds hold at constant rate, where the Johnson radius and the list-decoding capacity $1-R$ are separated by a constant. Our result also addresses a question left open by the recent work of Brakensiek, Chen, Putterman, Zhang, and Zheng~\cite{BCPZZ26}, who established capacity-approaching list-decodability for arbitrary evaluation points over sufficiently large \emph{prime} fields, leaving open whether analogous guarantees hold over general finite fields. We show that such guarantees cannot hold in general over extension fields of fixed characteristic.

\begin{theorem}[Not all RS codes are list-decodable beyond Johnson radius]
\label{thm:near-johnson-intro}
Let $q$ be any prime power, and let $R$ be either $q^{-2}$ or $(1-q^{-1})^2$. For every integer $t \ge 1$, let $n = q^{4t}$. Then there is an explicit received word $y \in \F_{q^{4t}}^{\,n}$ such that the full-length Reed--Solomon code over $\F_{q^{4t}}$ of rate $R$ has $n^{\Omega_q(\log n)}$ codewords at relative distance $1-\sqrt{R}+O_q(n^{-1/4})$ from $y$.
\end{theorem}

Together with the Johnson bound,~\cref{thm:near-johnson-intro} (proved in~\cref{sec:main-proof}) shows that the Johnson radius is essentially the exact threshold for polynomially bounded lists in these full-length Reed--Solomon codes. For $q$ ranging over prime powers, this phenomenon occurs at two infinite families of constant rates, $R = q^{-2}$ and $R = (1-q^{-1})^2$, which approach $0$ and $1$, respectively, as $q$ grows. At each of these rates, one can be $o(1)$ close to the Johnson radius and still get super-polynomially large lists. Whether the same phenomenon occurs at every fixed rate $R \in (0,1)$ remains open.

Thus, for a Reed--Solomon code of rate $R$ over a field of small characteristic to be $(1-\sqrt{R}+\eps,L)$-list-decodable, for a fixed $\eps > 0$ and with $L$ polynomial in $n$, one has to place restrictions on its set of evaluation points. Indeed, the fields $\F_{q^{4t}}$ in \cref{thm:near-johnson-intro} all have the characteristic of the fixed base field $\F_q$, and the codes in the theorem use every element of these fields as an evaluation point. In a companion paper~\cite{AGG26-companion}, we show a positive result in this regard: for a careful choice of \emph{explicit} evaluation points in an extension field, one can list-decode up to capacity, and in fact achieve a list size that is optimal as a function of the gap to capacity.

\medskip \noindent \textbf{Prime field RS codes.}
For prime fields, as discussed earlier, a striking recent result of Brakensiek, Chen, Putterman, Zhang, and Zheng~\cite{BCPZZ26} established that Reed--Solomon codes over sufficiently large prime fields can be list-decoded up to capacity, that is, up to a relative radius approaching $1-R$, for \emph{every} choice of evaluation points.  More precisely, for every constant rate $R$ and every $\eps > 0$, provided the prime is at least a suitable constant multiple (depending on $R$ and $\eps$) of the code length, every Reed--Solomon code of rate $R$ over the prime field is $(1-R-\eps,L)$-list-decodable with $L$ polynomial in the field size, and the list can be found in time polynomial in the field size~\cite[Corollary~5.1]{BCPZZ26}. The formulation in Harsha, Kumar and Saptharishi~\cite[Theorem~1.1]{HKS26} (and also the immediate follow-up~\cite{Jer26} to \cite{BCPZZ26}) requires only $p \geq n$ over a prime field $\F_p$, and therefore covers \emph{every} prime-field Reed--Solomon code of constant rate $R < 1$, including the full-length code (they also make the list size polynomial in the block length as opposed to field size).

The result of~\cite{BCPZZ26}, however, yields a polynomial list-size bound with a large exponent that depends doubly exponentially on $1/\eps$. It has been conjectured~\cite[Conjecture~1]{CGHLLPS26} that, over prime fields, list-decoding capacity can in fact be achieved with a constant list size $L=L(R,\eps)$. In the opposite direction, Krachun, Kazanin, and Hab{\"o}ck~\cite[Theorem~1]{KKH26} constructed arbitrarily large polynomial lists for Reed--Solomon codes over prime fields whose evaluation points form a multiplicative subgroup, but only at a vanishing gap of order $1/\log n$ below capacity.

Our second result refutes this conjecture definitively by exhibiting lists of polynomial size for full-length Reed--Solomon codes at a constant gap below capacity. Specifically, for every integer $m \geq 2$ with $m \equiv 2 \pmod 3$, it gives full-length Reed--Solomon codes over prime fields of rate $R \approx 2/(3m)$ with at least $\text{poly}(n)$ codewords in a Hamming ball of relative radius about $1-1/m$, which is about $1/(3m)$ below the capacity $1-R$.

\begin{restatable}[Linear-sized lists below capacity over prime fields]{theorem}{polylists}
\label{thm:poly-lists}
Let $m\ge 2$ be an integer such that $m\equiv 2\pmod{3}$. Then, there exist infinitely many primes $p$ such that the full-length Reed--Solomon code over $\F_p$ of length $n=p$ and dimension $\frac{2(p+m-1)}{3m}$ has an explicit received word $y \in \F_p^{\,n}$ at Hamming distance exactly $p-\frac{p-1}{m}$ from at least $n$ codewords.
\end{restatable}

We prove \cref{thm:poly-lists} in~\cref{sec:multiplicative-proof}. There is no technical dependence between Theorems~\ref{thm:near-johnson-intro} and~\ref{thm:poly-lists}, so one can read~\cref{sec:multiplicative-proof} independently from~\cref{sec:main-proof}.

Setting $m=2$ in~\cref{thm:poly-lists} gives us the following corollary.

\begin{corollary}[Linear lists below capacity over prime fields]
\label{thm:linear-lists}
\label{prop:multiplicative-warmup}
For infinitely many primes $p$, the full-length Reed--Solomon code over $\F_p$ of length $n = p$ and dimension $(p+1)/3$ has an explicit received word $y \in \F_p^{\,n}$ at Hamming distance exactly $(p+1)/2$ from at least $n$ codewords.
\end{corollary}

Via a simple list-amplification technique described in the overview below, one can boost the list size in~\cref{thm:poly-lists} to obtain any polynomial list size, at the expense of a larger decoding radius. This is the content of the following result.

\begin{restatable}[Arbitrary polynomial-lists]{theorem}{largelists}
\label{thm:large-lists}
    For any integer $\ell\ge 2$, there are infinitely many primes $p$ such that the full-length Reed--Solomon code over $\F_p$ of length $n = p$ and rate $R = \Omega(1/\ell)$ is not $(1-R-\Omega(1/\ell^2), \binom{n}{\ell}/2)$-list-decodable.
\end{restatable}

\paragraph{Organization.} \cref{sec:proof-ideas} provides an overview of the ideas behind the proofs of both cases. 
In \cref{sec:main-proof}, we prove the limitations of list-decoding at the Johnson bound for small characteristic fields (\cref{thm:near-johnson-intro}).
In \cref{sec:multiplicative-proof}, we prove the limitations over prime fields (\cref{thm:poly-lists} and  \Cref{thm:large-lists}). There is no technical dependence between Theorems~\ref{thm:near-johnson-intro} and \ref{thm:poly-lists} (with the latter being simpler), and so they can be read independently.  

\paragraph{Concurrent work.}
In independent recent works,
Dao, Kominers, Thaler and Zheng~\cite{DKTZ26} and Zhu~\cite{Zhu26} also prove a lower bound for list decoding full-length Reed--Solomon codes of rate $q^{-2}$ whenever $q$ is a power of two, with list size $n^{\Omega_q(\log n)}$ at relative radius $1-q^{-1}+o(1)$. Their results thus include the low-rate conclusion of \cref{thm:near-johnson-intro} when $q$ is a power of two. Neither of these works has any result concerning limitations of list decoding prime field Reed--Solomon codes, and in particular do not include the analogs of Theorems~\ref{thm:poly-lists} and \ref{thm:large-lists}.

\section{Proof overview}
\label{sec:proof-ideas}

To demonstrate a list-size lower bound of $L$ at relative radius
$\rho$, we need to exhibit a received word with $L$ distinct codewords
within relative Hamming distance $\rho$. Concretely, let $\F$ be a
finite field, set $n\coloneqq\abs{\F}$, and let $C\subseteq\F^n$ be
the full-length Reed--Solomon code of rate $R\in(0,1)$, evaluated at
all elements of $\F$. By interpolation, any received word
$y\in\F^n$ can be written as $y=(g(\alpha))_{\alpha\in\F}$ for a
polynomial $g\in\F[X]$ of degree less than $n$. Each codeword
$c_i\in C$ is the evaluation $c_i=(f_i(\alpha))_{\alpha\in\F}$ of a
message $f_i\in\F[X]$ of degree less than $Rn$. Thus $c_i$ agrees
with $y$ at a position $\alpha\in\F$ precisely when
$g(\alpha)-f_i(\alpha)=0$. Defining the \emph{error polynomial}
$p_i\coloneqq g-f_i$, we can therefore reformulate the problem as
finding distinct polynomials $p_1,\ldots,p_L\in\F[X]$ satisfying the
following two properties.
\begin{enumerate}
\item\label{item:error-roots}
Each $p_i$ has at least $(1-\rho)n$ distinct roots in $\F$.

\item\label{item:error-top}
The polynomials share all coefficients of degree at least $Rn$;
equivalently, $\deg(p_i-p_j)<Rn$ for every $i,j$.
\end{enumerate}
Indeed, the first property expresses the required agreements, and the second
follows from $p_i-p_j=f_j-f_i$. Conversely, given such polynomials, let
$g$ be their common high-degree part. Then $f_i\coloneqq g-p_i$ are
distinct messages of degree less than $Rn$, whose evaluations agree
with the evaluation of $g$ at the roots of $p_i$. Since evaluation is
injective on polynomials of degree less than $n$, this gives $L$
distinct codewords within distance $\rho n$ of a common received word.

The difficulty is to satisfy both properties simultaneously. On the
one hand, we can easily satisfy property~\ref{item:error-top} by taking
a common monomial $X^D$, with $D\geq Rn$, and adding a few terms of
degree less than $Rn$. However, this offers no control over the number
of roots in $\F$. On the other hand, we can choose arbitrary sets
$A_1,\ldots,A_L\subseteq\F$ of size at least $(1-\rho)n$ and take
$p_i\coloneqq\prod_{a\in A_i}(X-a)$. These polynomials automatically
satisfy property~\ref{item:error-roots}, but ensuring that their
high-degree coefficients agree is much more difficult.

Indeed, a root polynomial of degree $(1-\rho)n$ has about
$(1-\rho-R)n$ high-degree coefficients, with roughly
$n^{(1-\rho-R)n}$ possible coefficient tuples. For a radius $\rho$
bounded a constant below the capacity $1-R$, this number far exceeds
the at most $2^n$ subsets of $\F$, so direct pigeonholing does not
guarantee a large family with a common high-degree part. The counting
argument of Justesen and H{\o}holdt~\cite{JH01}, as reformulated
in~\cite{BKR10}, follows this approach, but applies only at radii
close to capacity.

This brings us to a natural question about polynomials over finite
fields: how does the sparsity of a polynomial relate to the number
of roots it has? Our constructions exploit this connection through
structured choices of the agreement sets $A_i$, allowing us to obtain
many roots while arranging the required cancellation of high-degree
terms.

In this overview, we reverse the order of the main proofs, beginning
with the simpler construction over prime fields before turning to
extension fields.

\subsection{Prime fields: choosing and moving the agreement sets}
\label{subsec:overview-prime}

Our prime-field construction starts with a set whose root polynomial is
especially sparse: a multiplicative subgroup $H\subseteq\F_p^\times$ of
size $t$, with root polynomial $X^t-1$. We obtain many agreement sets
$A_i$ by moving $H$ under low-degree maps. The key observation is that
the multiplicative symmetry of $H$ forces a long gap below the leading
term even after its roots have been moved. Specifically, if $f(0)=0$ and
$1\leq\deg(f)=d<t$, then
\[
  P_f(X)\coloneqq\prod_{u\in H}(X-f(u)),
  \qquad
  \deg(P_f-X^t)\leq t-\ceil{t/d}.
\]
Thus, low-degree maps preserve a common high-degree part, while maps
that are injective on $H$ preserve the number of distinct roots. This
separates the two requirements on our error polynomials
(\cref{lem:multiplicative-cancellation}).

For primes $p\equiv2\pmod3$, a convenient family is given by the shifted cubes $f_a(X)=(X+a)^3-a^3$, indexed by $a\in\F_p$. These maps are permutations of $\F_p$, so each agreement set $A_a=f_a(H)$ has size $t$. Choosing a proper subgroup with $t>3$ and $t\equiv2\pmod3$ also makes these sets pairwise distinct: a single moment of $A_a$ recovers $a$. Consequently, the $p$ error polynomials $P_{f_a}$ all have $t$ roots and share the top part $X^t$, with the remaining terms of degree at most about $2t/3$. Taking $t$ to be a fixed positive fraction of $p$ gives linear-sized lists at a constant gap below capacity, proving \cref{thm:poly-lists}. For example, taking $H$ to be the squares gives rate about $1/3$ and relative distance about $1/2$.

To obtain larger lists, we multiply $\ell$ of these error polynomials. Their root sets become unions of $\ell$ moved copies of $H$, and the products retain a common top part $X^{\ell t}$ with a gap of about $t/3$ below it. Choosing $H$ sufficiently small keeps the average overlap between the copies low, so many unions contain nearly $\ell t$ points. Now, we show that sufficiently many moments show that different choices of factors give distinct products. This yields lists of size $\Omega_\ell(n^\ell)$ at rate $\Theta(1/\ell)$ and a gap $\Omega(1/\ell^2)$ below capacity, as in \cref{thm:large-lists}. The details are in \cref{sec:multiplicative-proof}.

\subsection{Extension fields: subspaces and the BKR construction}
\label{subsec:overview-extension}
\label{subsec:overview-linear}

Over $\F_{q^m}$, we can instead exploit its vector space structure over
$\F_q$. The construction of Ben-Sasson, Kopparty and Radhakrishnan~\cite{BKR10}
chooses the agreement sets $A_i$ to be $\F_q$-linear subspaces. If $A$ has
dimension $r$, its \emph{subspace polynomial} is linearized:
\[
  p_A(X)\coloneqq\prod_{a\in A}(X-a)
  =X^{q^r}+\sum_{j=0}^{r-1}c_jX^{q^j}.
\]
The $q^r$ roots give the agreements, and the restriction to powers of
$q$ leaves very few high-degree coefficients to control. 

A useful illustration is offered by hyperplanes. Let
$\Tr:\F_{q^m}\to\F_q$ be the field trace. Every $\F_q$-linear functional on $\F_{q^m}$ is uniquely of the form
$x\mapsto\Tr(\lambda x)$. Thus every hyperplane is the kernel of such a
functional with $\lambda\neq0$, and its monic root polynomial is
$p_A=\lambda^{-n/q}\Tr(\lambda X)$, where $n = q^m$. Its leading term is $X^{n/q}$ and
its next possible term has degree $n/q^2$. Hence all
$(n-1)/(q-1)$ hyperplanes already give error polynomials with a common
top part for code dimension $n/q^2+1$, each with $n/q$ roots. This gives
a linear-sized list at a radius approaching the Johnson radius for
rate $q^{-2}$.

Allowing subspaces of other dimensions gives the more general BKR
construction. Its counting argument yields super-polynomial lists when
the rate tends to zero, but only polynomial lists at constant rate.
To obtain super-polynomial lists at constant rate, we need a much larger
family of agreement sets while retaining control of the high-degree
coefficients.

\subsection{From subspaces to quadratic agreement sets for super-polynomial lists}
\label{subsec:overview-quadratic}

For fixed $q$, a subspace containing a constant fraction of $\F_{q^m}$
has bounded codimension, leaving only polynomially many choices.
A natural next step is to replace linear equations by quadratic ones.
Their zero sets can still contain about $n/q$ points, where $n=q^m$,
but there are $q^{\Theta(m^2)}$ quadratic forms. This offers enough
agreement sets for a super-polynomial list at constant rate, provided
we can obtain error polynomials with a common high-degree part.

Here we encounter an additional issue. Although quadratic in the
$\F_q$-coordinates, $Q$ has much larger degree as a univariate
polynomial. Unlike a subspace polynomial, it can have repeated roots,
so its degree need not equal the size of its zero set. 
It turns out that understanding
these extra multiplicities will also help us align the high-degree
coefficients.

The link comes from the bilinear form
$B_Q(x,y)=Q(x+y)-Q(x)-Q(y)$. For each fixed $y$, this is a linear
functional of $x$, so the trace representation used in the hyperplane
example lets us write $B_Q(x,y)=\Tr(xL_Q(y))$ for a unique
coefficient $L_Q(y)$. Bilinearity makes $y\mapsto L_Q(y)$ an
$\F_q$-linear map, represented by a linearized polynomial of degree
less than $n$. The construction below will ensure $\deg(Q)<2n/q$.
Under this condition, we have the key identity
\begin{equation}
\label{eq:key-identity}
Q^q-Q\equiv(X^n-X)L_Q.
\end{equation}
The factor $X^n-X$ is natural: since $Q$ is $\F_q$-valued,
$Q^q-Q$ vanishes at every field element. The identity says more by
identifying the quotient with $L_Q$. For suitable $Q$, this factor
accounts for the extra root multiplicities, and removing part of it
will give the desired common high-degree part. We therefore seek a
large family in which these multiplicities can be controlled.

The hyperplane example suggests how to build such a family. There we
composed the trace with the linear maps $x\mapsto\lambda x$.
We now compose a fixed quadratic form $H$ with a richer collection of
linear maps $P$, giving $Q(x)=H(P(x))$. Its zero set is the preimage
under $P$ of the zero set of $H$. This preserves the quadratic
structure while allowing many agreement sets. Taking $P$ to be a
low-degree linearized polynomial also controls the univariate exponents.

A convenient choice of $H$ comes from the field structure itself.
For $m=4t$, take the norm $x\mapsto x\cdot x^{q^{2t}}$ from
$\F_{q^{4t}}$ to $\F_{q^{2t}}$, followed by the trace from
$\F_{q^{2t}}$ to $\F_q$. The norm multiplies two $\F_q$-linear
maps, and the trace makes the resulting quadratic form $\F_q$-valued.
We want every root of $L_Q$ to be a zero of $Q$. To arrange this,
we first choose a $t$-dimensional subspace $U$ on which $H$ vanishes.
On $U^\perp$, with orthogonality defined by the bilinear form of $H$,
we have $H(x+u)=H(x)$ for every $x \in U^\perp$ and $u\in U$. This suggests choosing
$P$ to have image $U^\perp$. Then $Q$ is unchanged by translation
along $P^{-1}(U)$ and vanishes on this subspace. 
In fact, \(\ker L_Q=P^{-1}(U)\), so \(Q\) vanishes on the entire kernel of \(L_Q\). This gives the roots at which we will later remove the extra multiplicity. 

We choose a monic $P$ of degree $q^t$, the smallest possible degree
for a polynomial whose kernel has $q^t$ points. This keeps the powers
of $q$ appearing in the two factors of the norm separated: the shift
by $2t$ in $P(X)^{q^{2t}}$ exceeds the range of indices in $P$.
Consequently, even after taking the trace, the polynomial representing
$Q$ on the field contains only mixed terms $X^{q^i+q^j}$ with
$i\neq j$. Its degree is therefore less than $2n/q$ (in fact at most $n/q+n/q^2$), as required for the key identity.

The same restriction controls the correction factor $L_Q$. Its powers
$X^{q^k}$ have indices between $t$ and $3t$, so it can be written
as $L_Q=S^{q^t}$ with $\deg(S)\leq q^{2t}$. Here the polynomial
root exists because every exponent is divisible by $q^t$ and every
coefficient has a $q^t$th root in the field. In particular, $L_Q$
has the polynomial $q$th root we will need.

The subspace $W=P^{-1}(U)$ has $q^{2t}$ points, all of which are
roots of $S$. These already exhaust its degree bound. Thus $S$ is
a scalar multiple of the subspace polynomial of $W$, and
$\deg(L_Q)=q^t\deg(S)=q^{3t}=n^{3/4}$. This explains the role
of the subspaces in our construction: they ensure that the correction
factor has all its roots among the zeros of $Q$, while its degree
remains small compared with $n$.

We can now return to the error polynomials. The extra input from
our construction is that every root of $L_Q$ lies in the field and
is a zero of $Q$. Using these two facts, the key identity \eqref{eq:key-identity} shows
that each such root has multiplicity in $Q$ exactly one greater
than its multiplicity in $L_Q$. Hence $L_Q$ divides $Q$, and
dividing by the polynomial $q$'th root established above gives a
polynomial $p_Q=Q/L_Q^{1/q}$ with precisely the same zeros as $Q$.

Once this division is justified, the common high-degree part follows
directly from the key identity. Indeed, dividing the key identity by $L_Q$ gives
$p_Q^q=X^n-X+Q/L_Q$, and rearranging we have
\[
  (p_Q-X^{n/q})^q\equiv Q/L_Q-X.
\]
Comparing degrees in the key identity gives
$q\deg(Q)=n+\deg(L_Q)$, so $\deg(Q/L_Q)<n/q$.
It follows that $p_Q=X^{n/q}-f_Q$ with $\deg(f_Q)<n/q^2$.
This partial removal of multiplicity preserves every zero and gives
error polynomials with the same leading term $X^{n/q}$, with all
remaining terms of degree less than $n/q^2$, exactly as we desired.

There are many choices of $U$: when selecting an ordered basis,
at least $q^{3t}-q^{2t}$ choices remain at each step while keeping
$H$ zero on the span. A fixed $Q$ accounts for at most $q^{2t^2}$
such bases, since the possible kernels of $P$ lie inside the fixed
$2t$-dimensional space $\ker(L_Q)$. Comparing these counts gives
at least $q^{t^2}/2=n^{\Omega_q(\log n)}$ distinct forms
(\cref{lem:norm-forms}).

Their zero sets have size
$\deg(Q/L_Q)=n/q-O_q(n^{3/4})$. At most $q-1$ forms give the
same codeword after normalization.
We therefore obtain a super-polynomial list at rate $R=q^{-2}$
with agreement fraction just below $\sqrt{R}$
(\cref{subsec:low-rate-construction}).

For the high-rate result, we use the agreement set
$\{\alpha:Q(\alpha)\neq1\}$, with slightly fewer than $(1-1/q)n$
points. Multiplying the factors $Q-c$ for $c\in\F_q\setminus\{1\}$
and again removing suitable multiplicities gives the required common
high-degree part at rate $R=(1-q^{-1})^2$, again just beyond the
Johnson radius (\cref{subsec:high-rate-construction}).

\section{Superpolynomial lists just past the Johnson radius in extension fields}
\label{sec:main-proof}

In this section, we prove \cref{thm:near-johnson-intro}. Throughout this section, we fix a prime power $q$ and an integer $t \geq 1$, and we set $m \coloneqq 4t$ to be the degree of the field extension $\F_{q^{4t}}/\F_q$ in which we work. For any integer $s \geq 1$, we write $\equiv$ to denote polynomial identities in $\F_{q^s}[X]$ and $=$ for equalities over $\F_{q^s}$. We call a polynomial $P \in \F_{q^s}[X]$ \textbf{$\F_q$-valued} if $P(\alpha) \in \F_q$ for every $\alpha \in \F_{q^s}$. The field \textbf{trace} from $\F_{q^s}$ to $\F_q$ is the polynomial $\Tr_{\F_{q^s}/\F_q} \in \F_{q^{4t}}[X]$ defined as
\begin{equation*}
\Tr_{\F_{q^s}/\F_q}(X) \coloneqq \sum_{i=0}^{s-1}{X^{q^i}}.
\end{equation*}
For the sake of brevity, we abbreviate $\Tr_{\F_{q^{4t}}/\F_q}$ to $\Tr$ throughout. Note that for any $\alpha \in \F_{q^{4t}}$, we have that $\Tr(\alpha)^q = \Tr(\alpha)$, and so $\Tr(\alpha) \in \F_q$. Thus, the trace map is an $\F_q$-valued polynomial.

Next, we recall linearized polynomials.

\begin{definition}[Linearized polynomials]
\label{def:linearized-polynomials}
A polynomial $L \in \F_{q^{4t}}[X]$ is said to be \textbf{linearized} if it has the form $L(X) \equiv \sum_{i=0}^{r}a_iX^{q^i}$ for some integer $r \geq 0$ and coefficients $a_0,\ldots,a_r \in \F_{q^{4t}}$. Such a polynomial is additionally said to be \textbf{symmetric} if $\Tr(L(\alpha)\beta) = \Tr(\alpha L(\beta))$ for every $\alpha,\beta \in \F_{q^{4t}}$.
\end{definition}

To clarify the nomenclature behind linearized polynomials, observe that, equivalently, a polynomial $L \in \F_{q^{4t}}[X]$ is linearized if it satisfies the identities $L(X+Y) \equiv L(X) + L(Y)$ and $L(cX) \equiv cL(X)$ for every $c \in \F_q$. Thus, when $\deg(L) < q^{4t}$, one can equivalently regard linearized polynomials as $\F_q$-linear maps $L : \F_{q^{4t}} \to \F_{q^{4t}}$. Consequently, the set of roots of $L$ over $\F_{q^{4t}}$ forms an $\F_q$-vector space. We call this set the \textbf{kernel} of $L$, denoted $\ker(L) \coloneqq \{\alpha \in \F_{q^{4t}} : L(\alpha) = 0\}$. Additionally, the \textbf{image} of $L$, denoted $\operatorname{im}(L) \coloneqq \{L(\alpha) : \alpha \in \F_{q^{4t}}\}$, also forms an $\F_q$-subspace.

Next, we state the following proposition, which characterizes $\F_q$-valued linearized polynomials.

\begin{prop}[$\F_q$-valued linearized polynomials]
\label{prop:fq-valued-linearized-polynomials}
All $\F_q$-valued linearized polynomials $L \in \F_{q^{4t}}[X]$ satisfying $\deg(L) < q^{4t}$ are of the form $\Tr(\lambda X)$ for some unique $\lambda \in \F_{q^{4t}}$.
\end{prop}
\begin{proof}
For every $\lambda \in \F_{q^{4t}}$, the polynomial $\Tr(\lambda X) \equiv \sum_{i=0}^{4t-1}{\lambda^{q^i}X^{q^i}}$ is an $\F_q$-valued linearized polynomial of degree less than $q^{4t}$. Moreover, the map $\lambda \mapsto \Tr(\lambda X)$ is an injective $\F_q$-linear map from $\F_{q^{4t}}$ into the $\F_q$-vector space of $\F_q$-valued linearized polynomials of degree less than $q^{4t}$. Each polynomial in this space is determined by its values on $\F_{q^{4t}}$, which form an $\F_q$-linear map from $\F_{q^{4t}}$ to $\F_q$, and hence the space has dimension at most $4t = \dim_{\F_q}(\F_{q^{4t}})$. Thus, the map $\lambda \mapsto \Tr(\lambda X)$ is a bijection, which gives both the existence and the uniqueness of $\lambda$.
\end{proof}

Next, we introduce quadratic forms, the class of polynomials of primary interest in this work.

\begin{definition}[Quadratic forms]
\label{def:quadratic-form}
A \textbf{quadratic form} is any $\F_q$-valued polynomial $Q \in \F_{q^{4t}}[X]$ of the form
\begin{equation*}
Q(X) \equiv \sum_{0 \leq i \leq j < 4t}{a_{i,j}X^{q^i+q^j}}
\quad\text{with } a_{i,j} \in \F_{q^{4t}}.
\end{equation*}
\end{definition}

To analyze quadratic forms, we consider their polarizations and the resulting linearized polynomial, which is the content of the following lemma.

\begin{lemma}[The polarization polynomial]
\label{lem:polarization}
Let $Q \in \F_{q^{4t}}[X]$ be any quadratic form. Then there is a unique linearized polynomial $L_Q \in \F_{q^{4t}}[X]$ with $\deg(L_Q) < q^{4t}$ satisfying
\begin{equation*}
Q(X+\alpha)-Q(X)-Q(\alpha) \equiv \Tr(XL_Q(\alpha))
\quad\text{ for all }\alpha \in \F_{q^{4t}}.
\end{equation*}
Moreover, the linearized polynomial $L_Q$ is symmetric.
\end{lemma}

\begin{proof}
Fix $\alpha \in \F_{q^{4t}}$, and define the polynomial
\begin{equation*}
    R_\alpha(X) \coloneqq Q(X+\alpha)-Q(X)-Q(\alpha).
\end{equation*}
First, observe that $R_\alpha(X)$ is a linearized polynomial with $\deg(R_\alpha) < q^{4t}$. Indeed, expanding things out, we see that
\begin{align*}
R_\alpha(X) & \equiv Q(X+\alpha)-Q(X)-Q(\alpha) \\
& \equiv \sum_{0 \leq i \leq j < 4t}{a_{i,j}\left((X+\alpha)^{q^i+q^j} - X^{q^i+q^j} - \alpha^{q^i+q^j}\right)} \\
& \equiv \sum_{0 \leq i \leq j < 4t}{a_{i,j}\left(\alpha^{q^j}X^{q^i} + \alpha^{q^i}X^{q^j}\right)}.
\end{align*}
Thus, $R_\alpha$ consists of monomials of the form $X^{q^i}$, implying that it is linearized. Moreover, by definition, because $Q$ is $\F_q$-valued, so is $R_\alpha$. Thus, by~\cref{prop:fq-valued-linearized-polynomials}, we can find a unique $\lambda(\alpha)$ satisfying $R_\alpha(X) \equiv \Tr(X\lambda(\alpha))$.

From the expansion above, observe that $R_{\alpha+\beta}(X) \equiv R_\alpha(X) + R_\beta(X)$. Thus, by uniqueness, we see that $\lambda(\alpha+\beta) = \lambda(\alpha) + \lambda(\beta)$ for any $\alpha,\beta \in \F_{q^{4t}}$. Similarly, $R_{c\alpha}(X) \equiv cR_\alpha(X)$ for every $c \in \F_q$ and $\alpha \in \F_{q^{4t}}$, so uniqueness gives $\lambda(c\alpha) = c\lambda(\alpha)$. Thus $\lambda$ is $\F_q$-linear. By interpolation, we can thus find a unique linearized polynomial $L_Q \in \F_{q^{4t}}[X]$ with $\deg(L_Q) < q^{4t}$ satisfying $L_Q(\alpha) = \lambda(\alpha)$ for all $\alpha \in \F_{q^{4t}}$. Moreover, its symmetry follows from the observation that $R_\alpha(\beta) = R_\beta(\alpha)$, which is a direct consequence of the definition of $R_\alpha$.
\end{proof}

As it turns out, a quadratic form $Q$ and its polarization polynomial $L_Q$ are connected through a distinct relationship, expressed by the following key identity.

\begin{lemma}[Key identity]
\label{lem:norm-factorization}
Let $Q$ be a quadratic form with $\deg(Q) < 2q^{4t-1}$, or equivalently, with the coefficient $a_{4t-1,4t-1}$ of the highest possible monomial $X^{2q^{4t-1}}$ equal to zero. Then
\begin{equation}
\label{eq:q-valued-identity}
Q^q-Q \equiv (X^{q^{4t}}-X)L_Q.
\end{equation}
\end{lemma}

\begin{proof}
Define the polynomial $G \coloneqq Q^q-Q-(X^{q^{4t}}-X)L_Q$. Because $\deg(Q) < 2q^{4t-1}$, we see that $\deg(G) < 2q^{4t}$. Moreover, from the fact that $Q$ is $\F_q$-valued, we see that $G(\alpha) = 0$ for all $\alpha \in \F_{q^{4t}}$. Thus, we can write $G \equiv (X^{q^{4t}}-X)H$ for $H \in \F_{q^{4t}}[X]$ satisfying $\deg(H) < q^{4t}$. Now, for every $\alpha \in \F_{q^{4t}}$, by~\cref{lem:polarization}, observe that
\begin{align*}
G(X+\alpha)-G(X) 
& \equiv (Q(X+\alpha) - Q(X))^q - (Q(X+\alpha) - Q(X)) - (X^{q^{4t}}-X)L_Q(\alpha) \\
& \equiv (\Tr(XL_Q(\alpha)) + Q(\alpha))^q- (\Tr(XL_Q(\alpha)) + Q(\alpha))
-(X^{q^{4t}}-X)L_Q(\alpha) \\
& \equiv \sum_{i=1}^{4t}{(XL_Q(\alpha))^{q^i}} - \sum_{i=0}^{4t-1}{(XL_Q(\alpha))^{q^i}} 
-(X^{q^{4t}}-X)L_Q(\alpha) \\
& \equiv (XL_Q(\alpha))^{q^{4t}}-XL_Q(\alpha)-(X^{q^{4t}}-X)L_Q(\alpha) \\
& \equiv 0.
\end{align*}
Since $G \equiv (X^{q^{4t}}-X)H$, we thus see that $H(X+\alpha) \equiv H(X)$ for all $\alpha \in \F_{q^{4t}}$. Combined with the bound $\deg(H) < q^{4t}$, we deduce that $H$ is a constant polynomial. Since every monomial in $G$ has degree at least $2$, we see that $H$ has no constant term. Hence $H \equiv 0$, which concludes the proof.
\end{proof}

Next, we will characterize the root structure of $Q$. To do so, we first distinguish the forms that vanish on the kernel of their polarization polynomial, which is an automatic feature when $q$ is odd, but not when $q$ is even.\footnote{To see this when $q$ is odd, notice that homogeneity of $Q$ yields $Q(2X) \equiv 4Q(X)$. Plugging in $X = \alpha$ in~\cref{lem:polarization}, we see that $Q(\alpha) = \tfrac{1}{2}\Tr(\alpha L_Q(\alpha))$, which directly implies properness.}

\begin{definition}[Proper quadratic forms]
\label{def:proper-quadratic-form}
A quadratic form $Q$ is said to be \textbf{proper} if $Q|_{\ker(L_Q)} = 0$, that is, $Q(\alpha) = 0$ for every $\alpha \in \ker(L_Q)$.
\end{definition}

With this backdrop, under mild assumptions on $Q$, we can now present a characterization of the root structure of $Q$, and more generally its level sets.

\begin{corollary}[Level sets]
\label{cor:level-sets}
Let $Q \in \F_{q^{4t}}[X]$ be a nonzero proper quadratic form with $\deg(Q) < 2q^{4t-1}$. If $L_Q$ splits over $\F_{q^{4t}}$, then $L_Q$ divides $Q$ and $\deg(Q) = (q^{4t}+\deg(L_Q))/q$. In this case, for every $c \in \F_q$,
\begin{equation*}
\abs{\{\alpha \in \F_{q^{4t}} : Q(\alpha) = c\}}
= \begin{cases}
\deg(Q)-\deg(L_Q) & \text{if }c = 0,\\
\deg(Q) & \text{if }c \neq 0.
\end{cases}
\end{equation*}
\end{corollary}

\begin{proof}
From the degree bound on $Q$, by \cref{lem:norm-factorization}, \cref{eq:q-valued-identity} holds. In particular, we see that $L_Q$ must divide $Q^q-Q \equiv \prod_{c \in \F_q}{(Q-c)}$. Since $Q$ is proper and $L_Q$ splits completely, we thus see that $\gcd(L_Q, Q-c) = 1$ for every nonzero $c \in \F_q$. Thus, we deduce that $L_Q$ divides $Q$.

Moreover, from \cref{eq:q-valued-identity}, observe that
\begin{equation*}
\frac{Q}{L_Q}\prod_{c \in \F_q^\times}(Q-c)
\equiv X^{q^{4t}}-X.
\end{equation*}
In particular, the polynomials $Q-c$ for $c \in \F_q \setminus \{0\}$ and $Q/L_Q$ all split completely and have no repeated roots. Furthermore, $Q/L_Q$ shares the same set of roots as $Q$. Comparing degrees thus gives $q\deg(Q)-\deg(L_Q) = q^{4t}$ and the stated counts.
\end{proof}

Finally, we show that there are super-polynomially many choices of quadratic forms satisfying the needed assumptions.

\begin{lemma}[A family of proper quadratic forms]
\label{lem:norm-forms}
There are at least $q^{t^2}/2$ distinct nonzero proper quadratic forms $Q \in \F_{q^{4t}}[X]$ with $\deg(Q) < 2q^{4t-1}$, for which $L_Q$ has degree $q^{3t}$ and all its roots lie in $\F_{q^{4t}}$, each with multiplicity $q^t$.
\end{lemma}

\begin{proof}
For any linearized polynomial $P \in \F_{q^{4t}}[X]$ with $\deg(P) \leq q^t$, let $Q$ be the remainder of $\Tr_{\F_{q^{2t}}/\F_q}(P(X)^{q^{2t}+1})$ modulo $X^{q^{4t}}-X$. First, observe that $Q$ is $\F_q$-valued, as the norm $z \mapsto z^{q^{2t}+1}$ maps $\F_{q^{4t}}$ into $\F_{q^{2t}}$. Moreover, by using $\deg(P) \leq q^t$, expanding $\Tr_{\F_{q^{2t}}/\F_q}(P(X)^{q^{2t}+1})$ and reducing modulo $X^{q^{4t}}-X$, one can check that $Q$ consists of monomials of the form $X^{q^i+q^j}$ for $0 \leq i < j < 4t$. Consequently, $Q$ is an $\F_q$-valued quadratic form satisfying the degree bound $\deg(Q) \leq q^{4t-1}+q^{4t-2} < 2q^{4t-1}$.

Now, for any $\alpha, \beta \in \F_{q^{4t}}$, from the linearity of $P$, observe that
\begin{align*}
\Tr(\alpha L_Q(\beta))
& = Q(\alpha+\beta)-Q(\alpha)-Q(\beta) \\
& = \Tr_{\F_{q^{2t}}/\F_q}(P(\alpha+\beta)^{q^{2t}+1}) - \Tr_{\F_{q^{2t}}/\F_q}(P(\alpha)^{q^{2t}+1}) - \Tr_{\F_{q^{2t}}/\F_q}(P(\beta)^{q^{2t}+1}) \\
& = \Tr_{\F_{q^{2t}}/\F_q}(P(\alpha)^{q^{2t}}P(\beta)) + \Tr_{\F_{q^{2t}}/\F_q}(P(\beta)^{q^{2t}}P(\alpha)) \\
& = \Tr(P(\alpha)^{q^{2t}}P(\beta)).
\end{align*}
By the bound $\deg(P) \leq q^t$, we can write $P(X) \equiv \sum_{i=0}^t{b_iX^{q^i}}$. Thus, by the linearity of trace and the identity $\Tr(\gamma^q) = \Tr(\gamma)$, we see that
\begin{align*}
\Tr(P(\alpha)^{q^{2t}}P(\beta)) 
& = \Tr\left(\left(\sum_{i=0}^t{b_i^{q^{2t}}\alpha^{q^{2t+i}}}\right)\left(\sum_{j=0}^t{b_j\beta^{q^j}}\right)\right) \\
& = \sum_{i=0}^t{\sum_{j=0}^t{\Tr\left(b_i^{q^{2t}}b_j\alpha^{q^{2t+i}}\beta^{q^j}\right)}} \\
& = \sum_{i=0}^t{\sum_{j=0}^t{\Tr\left(b_i^{q^{4t-i}}b_j^{q^{2t-i}}\alpha \beta^{q^{2t+j-i}}\right)}} \\
& = \Tr\left(\alpha \left(\sum_{i=0}^t{\sum_{j=0}^t{b_i^{q^{4t-i}}b_j^{q^{2t-i}}\beta^{q^{2t+j-i}}}}\right)\right).
\end{align*}
By the uniqueness of $L_Q$ from~\cref{lem:polarization}, we see that
\begin{equation*}
L_Q(X) \equiv \sum_{i=0}^t{\sum_{j=0}^t{b_i^{q^{4t-i}}b_j^{q^{2t-i}}X^{q^{2t+j-i}}}}.
\end{equation*}
In particular, we see that $\deg(L_Q) \leq q^{3t}$ and $L_Q$ is a $q^t$-th power.

It thus remains to show the properness of $Q$ and the complete splitting of $L_Q$, which will depend on further properties of $P$, and to count the resulting quadratic forms. To this end, let $Q_0(z) \coloneqq \Tr_{\F_{q^{2t}}/\F_q}(z^{q^{2t}+1})$ and $\langle u,v \rangle \coloneqq \Tr(u^{q^{2t}}v)$ for $u,v,z \in \F_{q^{4t}}$. Since the reduction modulo $X^{q^{4t}}-X$ does not change values on $\F_{q^{4t}}$, we have $Q(\alpha) = Q_0(P(\alpha))$ and $\Tr(\alpha L_Q(\beta)) = \langle P(\alpha),P(\beta) \rangle$ for every $\alpha,\beta \in \F_{q^{4t}}$. Moreover, from the polarization identity
\begin{equation*}
Q_0(u+v)-Q_0(u)-Q_0(v) = \langle u,v \rangle
\quad\text{for any } u,v \in \F_{q^{4t}},
\end{equation*}
we see that the $\F_q$-bilinear form $\langle \cdot,\cdot \rangle$ is symmetric. It is also nondegenerate, because $\Tr(\lambda X)$ is a nonzero polynomial of degree less than $q^{4t}$ for every $\lambda \neq 0$. For an $\F_q$-subspace $V \subseteq \F_{q^{4t}}$, we write
\begin{equation*}
V^\perp \coloneqq \{\alpha \in \F_{q^{4t}} : \langle \alpha, v \rangle = 0 \text{ for every } v \in V\}.
\end{equation*}
We will show that $Q$ has all the remaining properties in the lemma whenever $P$ satisfies the following two conditions.

\begin{enumerate}[label=(C\arabic*)]
\item\label{item:P-roots} The polynomial $P$ has $q^t$ roots in $\F_{q^{4t}}$. In particular, $P$ fully splits.
\item\label{item:P-isotropic} The function $Q_0$ vanishes on the subspace $U \coloneqq (\operatorname{im}(P))^\perp$.
\end{enumerate}

Indeed, suppose that $P$ satisfies~\ref{item:P-roots} and~\ref{item:P-isotropic}, and let us first show the full splitting of $L_Q$. Let $\beta \in \F_{q^{4t}}$. Observe that $L_Q(\beta) = 0$ exactly when $\Tr(\alpha L_Q(\beta)) = 0$ for every $\alpha \in \F_{q^{4t}}$. As $\Tr(\alpha L_Q(\beta)) = \langle P(\alpha),P(\beta) \rangle$, this happens exactly when $P(\beta)$ is orthogonal to every element of $\operatorname{im}(P)$, that is, when $P(\beta) \in (\operatorname{im}(P))^\perp = U$. Thus, we see that $\ker(L_Q) = P^{-1}(U)$.

Next, we count the elements of this kernel. Since $\langle \cdot,\cdot \rangle$ is nondegenerate, every $\F_q$-subspace $V \subseteq \F_{q^{4t}}$ satisfies $\dim(V^\perp) = 4t-\dim(V)$ and $(V^\perp)^\perp = V$. By~\ref{item:P-roots}, we have $\dim(\ker(P)) = t$, and hence $\dim(\operatorname{im}(P)) = 3t$ and $\dim(U) = 4t-3t = t$. Moreover, $U$ lies in the image of $P$. Indeed, for all $u,v \in U$, condition~\ref{item:P-isotropic} and the polarization identity of $Q_0$ give $\langle u,v \rangle = Q_0(u+v)-Q_0(u)-Q_0(v) = 0$. Thus $U \subseteq U^\perp = \operatorname{im}(P)$. Since $U \subseteq \operatorname{im}(P)$, the preimage $P^{-1}(U)$ consists of $\abs{U} = q^t$ cosets of $\ker(P)$, and hence $\abs{\ker(L_Q)} = q^t \cdot q^t = q^{2t}$. Since $L_Q$ is a $q^t$-th power, this shows the full splitting of $L_Q$.

We now show that $Q$ is proper, that is, that $Q(\beta) = 0$ for every $\beta \in \ker(L_Q)$. Let $\beta \in \ker(L_Q)$. Since $\ker(L_Q) = P^{-1}(U)$, the element $P(\beta)$ lies in $U$. By~\ref{item:P-isotropic}, the function $Q_0$ vanishes on $U$, and hence $Q(\beta) = Q_0(P(\beta)) = 0$. Thus $Q$ vanishes on $\ker(L_Q)$.

It thus remains to construct enough polynomials $P$ satisfying~\ref{item:P-roots} and~\ref{item:P-isotropic}. Let $\mathcal{U}$ be the set of $t$-dimensional $\F_q$-subspaces $U \subseteq \F_{q^{4t}}$ on which $Q_0$ vanishes. The following claim shows that each subspace in $\mathcal{U}$ gives exactly one monic polynomial $P$ satisfying~\ref{item:P-roots} and~\ref{item:P-isotropic}.

\begin{claim}[From subspaces to polynomials]
\label{claim:norm-forms-polynomial}
For every $U \in \mathcal{U}$, there is exactly one monic linearized polynomial $P_U \in \F_{q^{4t}}[X]$ of degree $q^t$ with $\operatorname{im}(P_U) = U^\perp$. Moreover, $P_U$ satisfies~\ref{item:P-roots} and~\ref{item:P-isotropic}.
\end{claim}

\begin{claimproof}
Fix bases $e_1,\ldots,e_{4t}$ of $\F_{q^{4t}}$ and $u_1,\ldots,u_t$ of $U$ over $\F_q$. The linearized polynomials $P(X) \equiv \sum_{i=0}^t{b_iX^{q^i}}$ form an $\F_q$-vector space of dimension $4t(t+1)$, and the condition $\operatorname{im}(P) \subseteq U^\perp$ amounts to the $4t^2$ equations $\langle P(e_k),u_l \rangle = 0$ for all $k \in [4t]$ and $l \in [t]$. Since $4t(t+1) > 4t^2$, some nonzero $P$ satisfies this condition. Because $\operatorname{im}(P) \subseteq U^\perp$, the kernel of $P$ has at least $q^{4t}/\abs{U^\perp} = q^t$ points, and since $\deg(P) \leq q^t$, it follows that $\deg(P) = q^t$ and $\abs{\ker(P)} = q^t$, and hence $\operatorname{im}(P) = U^\perp$. For $\lambda \in \F_{q^{4t}}$ defined by the equation $\lambda^{q^t} = b_t^{-1}$, observe that the polynomial $P_U(X) \coloneqq P(\lambda X)$ is monic with the same image and degree, which establishes the existence of such a polynomial.

For uniqueness, let $P'$ be another monic linearized polynomial of degree $q^t$ with $\operatorname{im}(P') = U^\perp$. Because both $P'$ and $P_U$ are monic, $P_U-P'$ is a linearized polynomial of degree less than $q^t$. Moreover, its image lies in $U^\perp$, which implies that its kernel has size at least $q^{4t}/\abs{U^\perp} = q^t$, which is impossible for a nonzero polynomial of degree less than $q^t$. Hence $P' \equiv P_U$. Finally, $P_U$ satisfies~\ref{item:P-roots} because $\abs{\ker(P_U)} = q^{4t}/\abs{U^\perp} = q^t$, and it satisfies~\ref{item:P-isotropic} because $(\operatorname{im}(P_U))^\perp = (U^\perp)^\perp = U$.
\end{claimproof}

Next, we show that $\mathcal{U}$ is large. Let $N_t \coloneqq \prod_{i=0}^{t-1}(q^t-q^i)$ denote the number of ordered bases of a $t$-dimensional $\F_q$-vector space.

\begin{claim}[Many subspaces]
\label{claim:norm-forms-subspaces}
We have $\abs{\mathcal{U}} \geq (q^{3t}-q^{2t})^t/N_t$.
\end{claim}

\begin{claimproof}
Since each $U \in \mathcal{U}$ has exactly $N_t$ ordered bases, it suffices to show that there are at least $(q^{3t}-q^{2t})^t$ tuples $(u_1,\ldots,u_t)$ of linearly independent vectors such that $Q_0$ vanishes on their span, which we will show by iteratively choosing each vector. Let $0 \leq i < t$, and suppose that $Q_0$ vanishes on the $\F_q$-span of linearly independent $u_1,\ldots,u_i$, which we denote by $U_i$. Let $\alpha \in \F_{q^{4t}}$. For all $c \in \F_q$ and $u \in U_i$, the polarization identity of $Q_0$ and its homogeneity show that $Q_0(c\alpha+u) = c^2Q_0(\alpha)+c\langle \alpha,u \rangle$. Hence $Q_0$ vanishes on the span of $U_i$ and $\alpha$ when $Q_0(\alpha) = 0$ and $\alpha \in U_i^\perp$. Thus, any choice of $\alpha \in Q_0^{-1}(0) \cap (U_i^\perp \setminus U_i)$ would suffice.

Next, we lower bound $\abs{Q_0^{-1}(0) \cap (U_i^\perp \setminus U_i)}$. First, observe that the norm map $z \mapsto z^{q^{2t}+1}$ maps $\F_{q^{4t}}^\times$ onto $\F_{q^{2t}}^\times$ and takes each value exactly $q^{2t}+1$ times. Moreover, the trace map $\Tr_{\F_{q^{2t}}/\F_q}$ has $q^{2t-1}-1$ zeros in $\F_{q^{2t}}^\times$. Thus, $Q_0$ has $1+(q^{2t}+1)(q^{2t-1}-1) = q^{4t-1}-(q-1)q^{2t-1}$ zeros in $\F_{q^{4t}}$. Now, observe that
\begin{equation*}
\abs{Q_0^{-1}(0) \cap (U_i^\perp \setminus U_i)} \geq \abs{Q_0^{-1}(0)} - \abs{Q_0^{-1}(0) \cap (\F_{q^{4t}} \setminus U_i^\perp)} - \abs{Q_0^{-1}(0) \cap U_i}.
\end{equation*}
Since $Q_0$ vanishes on $U_i$, we thus see that $\abs{Q_0^{-1}(0) \cap U_i} = \abs{U_i} = q^i$. As for $\abs{Q_0^{-1}(0) \cap (\F_{q^{4t}} \setminus U_i^\perp)}$, first observe that $U_i \subseteq U_i^\perp$, because $\langle u,v \rangle = Q_0(u+v)-Q_0(u)-Q_0(v) = 0$ for all $u,v \in U_i$. Hence the subspace $U_i^\perp$ is a union of cosets of $U_i$, and therefore so is $\F_{q^{4t}} \setminus U_i^\perp$. Now, fix a coset $\alpha+U_i$ with $\alpha \notin U_i^\perp$. For every $u \in U_i$, we have $Q_0(u) = 0$, and so the polarization identity of $Q_0$ gives $Q_0(\alpha+u) = Q_0(\alpha)+\langle \alpha,u \rangle$. Since $\alpha \notin U_i^\perp$, the map $u \mapsto \langle \alpha,u \rangle$ is a nonzero $\F_q$-linear functional on $U_i$, and hence it takes every value in $\F_q$ equally often. In particular, $Q_0(\alpha+u) = 0$ for exactly a $1/q$ fraction of the elements $u \in U_i$. Thus $\abs{Q_0^{-1}(0) \cap (\alpha+U_i)} = q^{i-1}$ for every $\alpha \notin U_i^\perp$. Summing over these cosets, we thus see that $\abs{Q_0^{-1}(0) \cap (\F_{q^{4t}} \setminus U_i^\perp)} = (q^{4t}-q^{4t-i})/q$. Plugging in the values for all three terms, we therefore find that
\begin{equation*}
\abs{Q_0^{-1}(0) \cap (U_i^\perp \setminus U_i)}
\geq (q^{4t-1}-(q-1)q^{2t-1})-\frac{q^{4t}-q^{4t-i}}{q}-q^i
= q^{4t-i-1}-(q-1)q^{2t-1}-q^i.
\end{equation*}
Since $i \leq t-1$, we have $q^{4t-i-1} \geq q^{3t}$ and $q^i \leq q^{2t-1}$, and hence this count is at least $q^{3t}-(q-1)q^{2t-1}-q^{2t-1} = q^{3t}-q^{2t}$. Multiplying these bounds over each $i \in \{0, 1, \ldots, t-1\}$ gives at least $(q^{3t}-q^{2t})^t$ such tuples, concluding the proof of this claim.
\end{claimproof}

Finally, we lower bound the number of resulting quadratic forms from such polynomials $P_U$ by upper bounding the number of subspaces $U$ that can give the same form $Q$. Indeed, suppose that $P_U$ gives the quadratic form $Q$. Then, by the identity $\Tr(\alpha L_Q(\beta)) = \langle P_U(\alpha),P_U(\beta) \rangle$ for every $\alpha,\beta \in \F_{q^{4t}}$, we see that $\ker(P_U)$ is a $t$-dimensional subspace of $\ker(L_Q)$. Moreover, since $P_U \equiv \prod_{\gamma \in \ker(P_U)}(X-\gamma)$, two different polynomials $P_U$ give rise to different kernels. Since $P_U$ determines $U = (\operatorname{im}(P_U))^\perp$, different subspaces $U$ give different polynomials $P_U$, and hence different kernels $\ker(P_U)$. Since each such kernel has exactly $N_t$ ordered bases, all among the $\abs{\ker(L_Q)}^t = q^{2t^2}$ $t$-tuples of elements of $\ker(L_Q)$, we therefore see that at most $q^{2t^2}/N_t$ different choices of $U$ give $Q$ as a quadratic form. Thus, by~\cref{claim:norm-forms-subspaces}, we conclude that there are at least
\begin{equation*}
\frac{\abs{\mathcal{U}}N_t}{q^{2t^2}}
\geq \frac{(q^{3t}-q^{2t})^t}{q^{2t^2}}
= q^{t^2}(1-q^{-t})^t
\geq \frac{q^{t^2}}{2}
\end{equation*}
different such quadratic forms $Q$. This concludes the proof of the lemma.
\end{proof}

\subsection{The low-rate construction}
\label{subsec:low-rate-construction}

In this subsection, we prove the low-rate part of \cref{thm:near-johnson-intro}.

\begin{proof}[Proof of \cref{thm:near-johnson-intro} at rate $q^{-2}$]
Set $n \coloneqq q^{4t}$ and $R \coloneqq q^{-2}$. Our received word will be the vector $y \coloneqq (\alpha^{n/q})_{\alpha \in \F_{q^{4t}}}$, i.e., the evaluation of the monomial $X^{n/q}$. For the family of at least $q^{t^2}/2$ quadratic forms $Q$ given by \cref{lem:norm-forms}, we will show that the corresponding vectors $(f(\alpha))_{\alpha \in \F_{q^{4t}}}$, where
\begin{equation*}
f \coloneqq X^{n/q}-\frac{Q}{L_Q^{1/q}},
\end{equation*}
are codewords near $y$. Indeed, first, by \cref{lem:norm-forms}, observe that $L_Q$ is a fully split polynomial that is a $q^t$-th power. Thus, we can write $L_Q^{1/q}$ to denote its unique polynomial $q$-th root. Moreover, by~\cref{cor:level-sets}, we see that $L_Q^{1/q}$ divides $Q$. Thus, $f$ is a well-defined function that is moreover a polynomial in $\F_{q^{4t}}[X]$. To see that $(f(\alpha))_{\alpha \in \F_{q^{4t}}}$ is a bona fide codeword, it remains to show that $f$ has degree less than $Rn$, which we will now show.

\begin{claim}[Degree bound]
\label{claim:low-rate-degree}
We have $\deg(f) < n/q^2 = Rn$.
\end{claim}

\begin{claimproof}
By~\cref{lem:norm-factorization}, observe that we have the following identity:
\begin{equation*}
f^q \equiv \left(X^{n/q} - \frac{Q}{L_Q^{1/q}}\right)^q \equiv X - \frac{Q}{L_Q}.
\end{equation*}
By \cref{lem:norm-forms,cor:level-sets}, the quotient $Q/L_Q$ has degree less than $n/q$. Hence, the right-hand side of the above identity has degree less than $n/q$. Comparing degrees gives $q\deg(f) < n/q$, which establishes this claim.
\end{claimproof}

Next, we show that $(f(\alpha))_{\alpha \in \F_{q^{4t}}}$ agrees with $y$ in at least $(\sqrt{R}-o(1))n$ coordinates.

\begin{claim}[Agreements near Johnson]
\label{claim:low-rate-agreements}
The codeword $(f(\alpha))_{\alpha \in \F_{q^{4t}}}$ agrees with $y$ in at least $n/q-q^{3t} = (\sqrt{R}-n^{-1/4})n$ positions.
\end{claim}

\begin{claimproof}
Since $y$ is the evaluation of the monomial $X^{n/q}$, the codeword $(f(\alpha))_{\alpha \in \F_{q^{4t}}}$ agrees with $y$ at exactly the roots in $\F_{q^{4t}}$ of the error polynomial $X^{n/q}-f \equiv Q/L_Q^{1/q}$. By~\cref{cor:level-sets}, the set of roots of $Q/L_Q^{1/q}$ is the same as that of $Q$ and has size $\deg(Q)-\deg(L_Q)$, which is at least $n/q-q^{3t}$ by the degree bound $\deg(L_Q) \leq q^{3t}$.
\end{claimproof}

Finally, observe that two quadratic forms $Q$ and $Q'$ in the family from~\cref{lem:norm-forms} yield the same codeword only when $Q' \equiv cQ$ for some $c \in \F_q^\times$. Indeed, since $\deg(f) < n$, the codeword determines $f$, and hence both $Q/L_Q^{1/q} \equiv X^{n/q}-f$ and $Q/L_Q \equiv X-f^q$, by the identity in the proof of Claim~1. Their ratio is $(L_Q^{1/q})^{q-1}$, so $L_Q^{1/q}$ is determined up to multiplication by an element of $\F_q^\times$, and consequently so is $Q$. Thus, we conclude that the full-length Reed--Solomon code over $\F_{q^{4t}}$ of rate $R$ has at least $q^{t^2}/(2(q-1)) = n^{\Omega_q(\log n)}$ codewords at relative distance $1-\sqrt{R}+O_q(n^{-1/4})$ from $y$, as \cref{thm:near-johnson-intro} asserts.
\end{proof}

\subsection{The high-rate construction}
\label{subsec:high-rate-construction}

In this subsection, we prove the high-rate part of \cref{thm:near-johnson-intro}.

\begin{proof}[Proof of \cref{thm:near-johnson-intro} at rate $(1-q^{-1})^2$]
Set $n \coloneqq q^{4t}$ and $R \coloneqq (1-q^{-1})^2$. Our received word will be the vector $y \coloneqq (\alpha^{(q-1)n/q})_{\alpha \in \F_{q^{4t}}}$, i.e., the evaluation of the monomial $X^{(q-1)n/q}$. For the family of at least $q^{t^2}/2$ quadratic forms $Q$ given by \cref{lem:norm-forms}, we will show that the corresponding vectors $(f(\alpha))_{\alpha \in \F_{q^{4t}}}$, where
\begin{equation*}
f \coloneqq X^{(q-1)n/q}-\frac{\sum_{j=1}^{q-1}Q^j}{(L_Q^{1/q})^{q-1}},
\end{equation*}
are codewords near $y$. Indeed, as in \cref{subsec:low-rate-construction}, $L_Q^{1/q}$ is a polynomial, and the denominator $(L_Q^{1/q})^{q-1}$ divides $L_Q$, which divides $Q$ by~\cref{cor:level-sets}. Thus, $f$ is a polynomial in $\F_{q^{4t}}[X]$. To see that $(f(\alpha))_{\alpha \in \F_{q^{4t}}}$ is a bona fide codeword, it remains to show that $f$ has degree less than $Rn$, which we will now show.

\begin{claim}[Degree bound]
We have $\deg(f) < (q-1)^2n/q^2 = Rn$.
\end{claim}

\begin{claimproof}
Let $E \coloneqq X^{(q-1)n/q}-f$ be the quotient in the definition of $f$. Adding and subtracting $(Q/L_Q^{1/q})^{q-1}$, write $f$ as
\begin{equation*}
f \equiv \left(X^{(q-1)n/q}-\left(\frac{Q}{L_Q^{1/q}}\right)^{q-1}\right)-\left(E-\left(\frac{Q}{L_Q^{1/q}}\right)^{q-1}\right).
\end{equation*}
We will show that both differences on the right-hand side have degree less than $Rn$.

For the first difference, by \cref{claim:low-rate-degree} in \cref{subsec:low-rate-construction}, we can write $Q/L_Q^{1/q} \equiv X^{n/q}-g$ with $\deg(g) < n/q^2$. Expanding $(X^{n/q}-g)^{q-1}$, every term other than $X^{(q-1)n/q}$ has degree at most $(q-2)n/q+\deg(g) < (q-2)n/q+n/q^2 = Rn$, and hence the first difference has degree less than $Rn$.

For the second difference, since $\sum_{j=1}^{q-1}Q^j \equiv (Q^q-Q)/(Q-1)$, by~\cref{lem:norm-factorization}, observe that we have the following identity:
\begin{equation}
\label{eq:high-rate-error-poly}
E \equiv \frac{X^{q^{4t}}-X}{Q-1}\,L_Q^{1/q}.
\end{equation}
By~\cref{cor:level-sets}, $Q-1$ has $\deg(Q)$ distinct simple roots in $\F_{q^{4t}}$, implying that $(X^{q^{4t}}-X)/(Q-1)$ is a polynomial. Multiplying the identity by $Q/L_Q^{1/q}$ and subtracting $(Q/L_Q^{1/q})^q \equiv X^{q^{4t}}-X+Q/L_Q$, which also follows from \cref{lem:norm-factorization}, gives
\begin{equation*}
\frac{Q}{L_Q^{1/q}}\left(E-\left(\frac{Q}{L_Q^{1/q}}\right)^{q-1}\right)
\equiv \frac{(X^{q^{4t}}-X)Q}{Q-1} - \frac{Q^q}{L_Q}
\equiv \frac{X^{q^{4t}}-X}{Q-1}-\frac{Q}{L_Q}.
\end{equation*}
By \cref{lem:norm-forms,cor:level-sets}, we have $\deg((X^{q^{4t}}-X)/(Q-1)) = n-\deg(Q) < (q-1)n/q$ and $\deg(Q/L_Q) < n/q$. Hence, the right-hand side of the above identity has degree less than $(q-1)n/q$. Since $Q/L_Q^{1/q} \equiv X^{n/q}-g$ has degree $n/q$, comparing degrees gives that the second difference has degree less than $(q-2)n/q < Rn$. Both differences thus have degree less than $Rn$, and therefore $\deg(f) < Rn$.
\end{claimproof}

Next, we show that $(f(\alpha))_{\alpha \in \F_{q^{4t}}}$ agrees with $y$ in at least $(\sqrt{R}-o(1))n$ coordinates.

\begin{claim}[Agreements near Johnson]
\label{claim:high-rate-agreements}
The codeword $(f(\alpha))_{\alpha \in \F_{q^{4t}}}$ agrees with $y$ in at least $(q-1)n/q-q^{3t-1} \geq (\sqrt{R}-n^{-1/4})n$ positions.
\end{claim}

\begin{claimproof}
Since $y$ is the evaluation of the monomial $X^{(q-1)n/q}$, the codeword $(f(\alpha))_{\alpha \in \F_{q^{4t}}}$ agrees with $y$ at exactly the roots in $\F_{q^{4t}}$ of the error polynomial $X^{(q-1)n/q}-f \equiv E$. By~\cref{eq:high-rate-error-poly} in the proof of the degree bound, the polynomial $(X^{q^{4t}}-X)/(Q-1)$ divides the error polynomial $E$. This quotient has roots exactly outside $Q^{-1}(1)$, and the remaining factor $L_Q^{1/q}$ has all its roots in $Q^{-1}(0)$ because $Q$ is proper. Thus, the codeword agrees with $y$ exactly outside the level set $Q^{-1}(1)$. By~\cref{cor:level-sets}, this level set has size $\deg(Q) = (n+\deg(L_Q))/q$, which is at most $n/q+q^{3t-1}$ by the degree bound $\deg(L_Q) \leq q^{3t}$.
\end{claimproof}

Finally, observe that two quadratic forms $Q$ and $Q'$ in the family from~\cref{lem:norm-forms} yield the same codeword only when $Q' \equiv Q$. Indeed, the codeword $(f(\alpha))_{\alpha \in \F_{q^{4t}}}$ disagrees with $y$ exactly on the level set $Q^{-1}(1)$, which by~\cref{cor:level-sets} consists of the $\deg(Q)$ roots of $Q-1$. These roots determine $Q-1$ up to a nonzero constant factor, and the constant term $Q(0)-1 = -1$ fixes that factor. Thus, we conclude that the full-length Reed--Solomon code over $\F_{q^{4t}}$ of rate $R$ has at least $q^{t^2}/2 = n^{\Omega_q(\log n)}$ codewords at relative distance at most $1-\sqrt{R}+O_q(n^{-1/4})$ from $y$, as \cref{thm:near-johnson-intro} asserts.
\end{proof}

\section{Growing lists below capacity over prime fields}
\label{sec:multiplicative-proof}

\subsection{Linear-sized lists from multiplicative subgroups}

Throughout this section, we fix a prime $p$ and set the block length to be $n=p$, with our main focus being the full Reed--Solomon code over $\F_p$. For integers $d, t$ with $1\leq d<t$ such that $t\mid p-1$, let $k=t-\ceil{t/d}+1$, $\gamma \in \F_p^\times$ be a generator of $\F_p^\times$, and let $H \coloneqq \{\gamma^{(p-1)/t}, \gamma^{2(p-1)/t}, \dots, \gamma^{p-1}\}$. Let
    \begin{equation*}
        P_f(X) \coloneqq \prod_{u \in H}(X-f(u)).
    \end{equation*}

\begin{lemma}[Coefficient cancellation on a multiplicative subgroup]
\label{lem:multiplicative-cancellation}
Let $f \in \F_p[X]$ have degree $d$, where $1 \leq d < t$ and $f(0) = 0$.  The coefficient of $X^{t-j}$ in $P_f$ vanishes for every integer $j$ with $1 \leq j < t/d$. Equivalently, $\deg(X^t-P_f) \leq t-\ceil{t/d}$.
\end{lemma}

\begin{proof}
Let $\omega = \gamma^{(p-1)/t}$ and write $f(X) \equiv \sum_{r=1}^d a_rX^r$. As a polynomial in the coefficients $a_r$, the coefficient of $X^{t-j}$ in $P_f$ is homogeneous of degree $j$ and invariant under $a_r \mapsto \omega^r a_r$, since this replaces $f(u)$ by $f(\omega u)$. Give $a_r$ weight $r$. Every monomial that occurs must have weight divisible by $t$, but its weight lies between $j$ and $dj$. There is no such monomial when $1 \leq j < t/d$, proving the lemma.
\end{proof}

\begin{lemma}
\label{lem:cubics-are-amaze}
Let $p\equiv 2 \pmod 3$ be a prime, and let $H\subseteq \F_p^\times$ be a multiplicative subgroup of size $t\equiv 2 \pmod 3$. Let $f_a(x) = (x+a)^3 - a^3$ for any $a\in \F_p$. For any $a\neq b\in \F_p$, $f_a(H)\neq f_b(H)$.
\end{lemma}
\begin{proof}
    Firstly, we note that since $p\equiv 2 \pmod{3}$, the functions $f_a$ are all injective. Now, We do a simple moment calculation to get the following:
    \begin{align*}
        \sum_{h\in H}f_a(h)^{(t+1)/3} & = \sum_{h\in H} h^{t+1}+\sum_{h\in H}(t+1)/3\cdot 3a \cdot h^t + \sum_{h\in H} p_a(h)\\
        & = a(t+1)\cdot |H| + \sum_{h\in H} (h+p_a(h))\\
        & = a(t+1)\cdot |H|
    \end{align*}
    where $p_a(h)$ is a polynomial in $h$ with no constant term and degree less than $t$. Note that $t+1<p$ since $H$ is not the full multiplicative group as $p-1\equiv 2 \not\equiv t\pmod 3$. Thus, we get that $f_a(H)$ and $f_b(H)$ are distinct.
\end{proof}

\begin{lemma}\label{lem:many-different-poly}
   Let $f\in \F_p[X]_{\leq d}$ be any polynomial such that $|f(H)|=|H|$, $f(0)=0$, and let $g \coloneqq X^t-P_f(X)$. Then $\deg(g) < k$ and \[d_H((\alpha^t)_{\alpha \in \F}, (g(\alpha))_{\alpha \in \F})=n-t.\] 
\end{lemma}

\begin{proof}
    Observe that the degree bound follows from \Cref{lem:multiplicative-cancellation} and for any $\alpha \in \F_p$, $\alpha^t - g(\alpha)=P_f(\alpha)$. Since $f(H)$ has size at least $|H|$, $P_f$ has at least $|H|$ roots, and these are exactly the evaluation points for which $\alpha^t = g(\alpha)$.
\end{proof}

We now have the necessary ingredients to establish \Cref{thm:poly-lists}. 

\polylists*
\begin{proof}
    Let $p\equiv 2 \pmod 3, p\equiv 1 \pmod{m}$ with $p\ge 3m+2$, let $t = \frac{p-1}{m}$, $d=3$. Observe that $H = \{\gamma^m,\gamma^{2m}, \dots, \gamma^{p-1}\}$. Thus, \[k=t-\frac{t+1}{3}+1 = \frac{2(t+1)}{3} = \frac{2(p+m-1)}{3m}.\] Note that $t\equiv 2 \pmod 3$ since $p-1 \equiv 1$ and $m\equiv 2\pmod 3$.

    For any $a\in \F_p$, consider $f_a(X)=(X+a)^3 -a^3$ and observe that all of them are injective on all of $\F_p$ and not just $H$ since $p\equiv 2\pmod 3$. It is now sufficient to prove that for any $a\ne b$, $f_a(H)\ne f_b(H)$ since then for the received word $y = \{\alpha^t\}_{\alpha \in \F_p}$, and for all $a\in \F_p$, we have that $\Delta(y, \{\alpha^{t}-P_{f_a}(\alpha)\}_{\alpha \in \F_p})=n-t$. This follows from \cref{lem:cubics-are-amaze} and \cref{lem:many-different-poly}.

    There are infinitely many primes which are $2\pmod 3$ and $1\pmod m$ since $3$ and $m$ are co-prime due to Dirichlet's theorem.
\end{proof}

\subsection{Polynomially growing lists via amplification}\label{subsec:amplification-poly-lists}

Let $\RS((\alpha_1, \dots, \alpha_n), k)$ be a Reed--Solomon code such that there is a polynomial $y: \F_q\to \F_q$ of degree $D\geq k-1$, and message polynomials $g_1, \dots, g_L$ such that \[d_H((y(\alpha_1), \dots, y(\alpha_n)), (g_i(\alpha_1), \dots, g_i(\alpha_n)))\leq n-a\]
then for any $S\subseteq [L]$ of size $\ell$, let \[G_S = y^{\ell}-\prod_{s\in S}(y-g_s)\] and let $Y = y^{\ell}$.

Observe that $\deg G_S \leq (\ell-1)D+k-1$, and they all agree with the received word $Y$ in $|\{i: \exists s\in S\mid g_s(\alpha_i)=y(\alpha_i)\}|$ which we can hope is as big as $\ell a$.

In particular, if the sets $Z_s = \{i\mid g_s(\alpha_i)=y(\alpha_i)\}$ were disjoint, we get that $\RS((\alpha_1, \dots, \alpha_n),D(\ell-1)+k)$ has a received word, and $\ge \binom{L}{\ell}$ codewords at distance $\leq n-a\ell$.

For a different construction, we may require a more fine-grained analysis to show that these $\binom{L}{\ell}$ codewords are distinct and many of the sets are big.

\largelists*
\begin{proof}
    We use the above amplification idea on the construction in \Cref{thm:poly-lists}. 

    For simplicity let $\eps = 1/\ell$. Let $m = 12\ell^2+2$, and $p$ be any sufficiently large prime such that $p\equiv 1 \pmod m$ and $p\equiv 2 \pmod 3$. Now, $t = (p-1)/m$ and $k \leq \left\lceil\frac{2(p+1)}{3m}\right\rceil$.

    Then, we know by \cref{thm:poly-lists} that if we let $H = \{\gamma^m, \gamma^{2m}, \dots, \gamma^{p-1}\}$, let $f_a(x)=(x+a)^3-a^3$, then $d_H((\alpha^t)_{\alpha \in \F_p}, (\alpha^t - P_{f_a}(\alpha))_{\alpha \in \F_p})=n-t$ and all $f_a(H)$ are distinct.

    For $S\subseteq \F_p$ of size $\ell$, let $P_S=\prod_{s\in S}P_{f_s}$, and let $Y=\{\alpha^{\ell t}\}_{\alpha \in \F_p}$. Let $G_S = X^{\ell t}-P_S(X)$. 
    Let $K = t(\ell-1)+k = O(p/\ell)$. Thus, the full Reed--Solomon code of rate $R = \frac{K}{p}=\Theta(1/\ell)=\Theta(\eps)$ contains the messages given by $\{G_S(\alpha)\}_{\alpha \in \F}$. 
    
    We also have that \[d_h((\alpha^{\ell t})_{\alpha \in \F_p}, (\alpha^{\ell t}-P_S(\alpha))_{\alpha \in \F_p}) = n -\left|\bigcup_{s\in S}f_s(H)\right|.\]

    Thus, it suffices to prove that for many $S\subseteq \F_p$ of size $\ell$, the resulting polynomials $G_S$ are distinct and $\left|\bigcup_{s\in S}f_s(H)\right|>K+\Omega(k)$.
\newcommand{\E}{\mathbb E}
    We prove the second part first. Observe that 
    \begin{align*}
        \E_{S\subseteq \F_p, |S|=\ell}\sum_{(s_1,s_2) \in \binom{S}2}\left|f_{s_1}(H)\cap f_{s_2}(H)\right|& \leq  \binom{\ell}{2}\E_{s_1 \in \F_p, s_2 \in \F_p\setminus \{s_1\}} |\{(u, v)\in H\times H\mid f_{s_1}(u)=f_{s_2}(v)\}|\\
        & \leq \binom{\ell}{2}\frac{\sum_{s_1\in \F_p, u, v\in H} \E_{s_2\in \F_p\setminus \{s_1\}}\mathbf{1}(f_{s_1}(u)=f_{s_2}(v))}{p}\\
        & \le \binom{\ell}{2}\cdot |H|^2\cdot \frac{2}{p-1} \quad \le \frac{\ell^2t^2}{p} \quad \le \frac{\ell t}{12\ell}\\
        & \le \frac{t}{12}
    \end{align*}
    The key observation is that for any fixed $s_1, u, v$, there are at most $2$ choices of $s_2$ such that $(u+s_1)^3-s_1^3 = (v+s_2)^3 - s_2^3$ since this is a quadratic equation in $s_2$.

    Thus, $|\{S\mid \sum_{(s_1,s_2) \in \binom{S}{2}}\left|f_{s_1}(H)\cap f_{s_2}(H)\right|\le  t/6\}|\ge \binom{p}{\ell}\cdot \frac{1}{2}$. In particular, this means that for at least $\frac{\binom{n}{\ell}}{2}$ different choices of $S$, we have that $\left|\bigcup_{s\in S} f_s(H)\right|\ge t\ell - \sum_{(s_1,s_2) \in \binom S2}\left|f_{s_1}(H)\cap f_{s_2}(H)\right|\ge t\ell - \frac{t}{6}=K+\Omega(k)$.

    Now, it suffices to prove that all these polynomials $P_S$ are distinct. Thus, we prove the following stronger lemma which is sufficient.
    \begin{lemma}
        Fix $\ell\ge 1, m\ge 2$ with $m\equiv 2 \pmod 3$. For every sufficiently large prime $p$ with $p\equiv 1 \pmod m$ and $p\equiv 2 \pmod 3$, for any $S_1\neq S_2 \subseteq \F_p$, with $|S_1|=|S_2|=\ell$, \[\prod_{s\in S_1}P_{f_{s}}(X)\neq \prod_{s\in S_2}P_{f_{s}}(X)\]
    \end{lemma}
    \begin{proof}
        As a variant on the computation for the linear lists where we showed that $\sum_{h\in H} f_a(h)^{(t+1)/3}$ uniquely recovers $a$, we repeat the same argument with higher moments. 
        
        Let \[m_{S, r}=\sum_{s\in S}\sum_{h\in H}f_{s}(h)^{(t+1)/3+r}\]

        Observe that if $P_{S_1}=P_{S_2}$ then $m_{S_1, r}=m_{S_2, r}$ for all $r\in \mathbb N$. Additionally, from \Cref{thm:poly-lists}, we have that $m_{S, 0}= (t+1)|H|\sum_{s\in S} s$.

        We now only use $r \in \{0,1,\ldots, 2\ell\}$. Similarly, \begin{align*}
        \sum_{h\in H}f_a(h)^{(t+1)/3+r} & = \sum_{j\ne t}C_{r, j}(a)\sum_{h\in H}h^j+\sum_{h\in H} C_{r, t}(a)h^{t}\\
        & = C_{r, t}(a)|H|
    \end{align*}
    Here we use that every exponent satisfies $j\leq t+1+3r < 2t$, since we will assume that $p$ is sufficiently large such that $t>6\ell+1$.

    Now, it suffices to prove that for any fixed $\ell, m, r$, for large enough choices of $\F_p$, $C_{r, t}(a)$ is exactly a degree $ 3r+1$ function of $a$. Specifically, $f_a(X)^{(t+1)/3+r}=(X^{3}+3aX^2+3a^2X)^{(t+1)/3+r}$. Observe that this is a homogeneous bivariate polynomial in variables $a, X$ of degree $3(r+(t+1)/3)=t+3r+1$. Thus, the $X^t$ variable will have a corresponding $a^{3r+1}$ term. Our goal is now to simply compute what this term is. Observe that $C_{r, t}(a)=C_{r, t}(1)a^{3r+1}$ due to the homogeneity. Thus, we only focus on proving $C_{r, t}(1)\neq 0\pmod{p}$ for sufficiently large primes $p$.

    Let $d_r = (t+1)/3+r$ and $e_r = 3r+1$. Now, observe that if we use $b$ terms of the form $3a^2X$, to get the term of degree $e_r$ in $a$, we must use $e_r-2b$ terms of the form $3aX^2$ and the remaining of the form $X^3$. Thus,
    \begin{align*}
 C_{r,t}(1)
 &=\left(
   \sum_{b=0}^{\lfloor e_r/2\rfloor}
   3^{e_r-b}\binom{d_r}{b,e_r-2b, d+b-e_r}
   \right)\\
 &=\left(
   \sum_{b=0}^{\lfloor e_r/2\rfloor}
   \frac{1}{b!(e_r-2b)!}
   \prod_{i=0}^{e_r-b-1}(3d_r-3i)
   \right)\\
 &=\left(
   \sum_{b=0}^{\lfloor e_r/2\rfloor}
   \frac{1}{b!(e_r-2b)!}
   \prod_{i=0}^{e_r-b-1}(t+e_r-3i)
   \right),
\end{align*}

Observe that we are only interested in $C_{r, t}\pmod p$ and $t \equiv - 1/m \pmod p$. Hence,
\[m^{e_r}\cdot e_r!\cdot C_{r, t}(1)=\left(\sum_{b=0}^{\lfloor e_r/2\rfloor}
   \frac{e_r!}{b!(e_r-2b)!}
   \prod_{i=0}^{e_r-b-1}\left(\frac{-1}{m}+e_r-3i\right)\cdot m^{e_r}
   \right)\pmod{p} \]

Thus, \[M_r=\sum_{b=0}^{\lfloor e_r/2\rfloor}
   \frac{e_r!}{b!(e_r-2b)!}
   \prod_{i=0}^{e_r-b-1}\left(\frac{-1}{m}+e_r-3i\right)\cdot m^{e_r}
   \equiv (-1)^{e_r}\pmod m.\] 
   
   Hence, picking any prime $p>|M_r|, (6\ell+1)m+1$ for all $r\in \{0,1, \ldots,  2\ell\}$, we get that $M_r\pmod p \neq 0$ and hence $C_{r, t}(1)\neq 0\pmod p$. Consequently, given $\prod_{s\in S} P_{f_s}(X)$, we can recover $\sum_{s\in S} s^{3r+1}$ for $r\in \{0,1, \dots, 2\ell\}$.

   We need to simply show that given these moments, we can recover the set $S$. Let $S_1, S_2$ be any two such sets for which all above moments are equal, and $T$ be their symmetric difference, i.e., $T=(S_1\cup S_2)\setminus (S_1\cap S_2)$. Now, for $x\in T$, let $c_x = (-1)^bx$ if $x\in S_b$. Since all the above moments are the same, we have that \[\sum_{x\in T}c_x (x^{3})^r=0\] for all $r \in \{0,1,\ldots, 2\ell\}$. Thus, either all $c_x =0 \ \forall x\in T \implies T\subseteq \{0\}$, or the resulting Vandermonde matrix of $\{(x^3)^r\}_{r\in \{0, \ldots,  |T|-1\}, t\in T}$ is singular. But, since $p\equiv 2 \pmod 3$, the cubing $x\mapsto x^3$ map is injective and the resulting Vandermonde matrix is invertible. Thus, $T \subseteq \{0\}$ but the size of $T$ is $2\ell - 2|S_1\cap S_2|$ which is even. Hence, for any distinct $S_1, S_2$, there are some unequal moments as desired and we get the desired result.
\end{proof}
   Now, we can simply take the messages $G_S$ for the $\geq \binom{n}{\ell}/2$ sets $S$ for which $\sum_{s_1, s_2 \in \binom{S}{2}}|f_{s_1}(H)\cap f_{s_2}(H)|\leq t/6$ and the result follows.
\end{proof}

\subsection*{Acknowledgments}
We thank Zihan Zhang for helpful discussions. We also thank the Simons Institute for the Theory of Computing for hosting us during the Fall 2026 program on \emph{Pseudorandomness \& High-Dimensional Expansion}.

\subsection*{AI methodology}

Initial versions of our key statements were established by GPT-6 Astra Pro over various conversations. Our main contributions are in abstracting and simplifying the proof ideas, and in the exposition. In particular, the proofs presented in the paper are substantially simpler and clearer in our opinion.

Immediately following the breakthrough result of \cite{BCPZZ26}, we wondered if their results can be improved, specifically whether beyond Johnson-radius list-decoding is possible for all Reed--Solomon codes, and whether list sizes can be improved to a fixed constant for large prime fields. 

In conversations with GPT-6 Astra Pro, we started exploring both ideas. Soon after, the AI gave us an early construction of $n^{\Omega(\log n)}$ list-size just beyond Johnson for small characteristic fields using a result of \cite{Sch20}. We closely understood  this construction and then worked extensively to simplify it over the next few weeks. In this paper, we present a completely elementary proof of the lower bound.

For the prime field case, GPT-6 similarly arrived at a non-elementary construction of low-degree functions after finding \cref{lem:multiplicative-cancellation} to prove the requisite lower bounds.  After understanding the approach, we first simplified the constructions by choosing the ``moving'' functions $f$ instead as \emph{random} low-degree polynomials. Then, with the help of AI, we further refined it to the simple choice of $(x+a)^3 - a^3$ presented in the paper (at the expense that the result as written now only works for infinitely many as opposed to all primes). 

The amplification argument of \cref{subsec:amplification-poly-lists} was then discovered in discussions with GPT-6 where we looked for generic ways to bootstrap the lower bounds.

We take full responsibility for the correctness and quality of exposition of the results presented in this paper.

\bibliographystyle{alpha}
\bibliography{ref-ld-lb}

@misc{AGG26-companion,
      title={Explicit Reed-Solomon Subspace Designs: Optimal List and Line Decoding}, 
      author={Omar Alrabiah and Rohan Goyal and Venkatesan Guruswami},
      year={2026},
      note = {\textit{arXiv preprint arXiv:2610.36215}},
      eprint={2610.36215},
      archivePrefix={arXiv},
      primaryClass={cs.IT},
      url={https://arxiv.org/abs/2610.36215}, 
}

@article{AGGLZ25,
  author  = {Omar Alrabiah and Zeyu Guo and Venkatesan Guruswami and Ray Li and Zihan Zhang},
  title   = {Random {Reed--Solomon} Codes Achieve List-Decoding Capacity with Linear-Sized Alphabets},
  journal = {Advances in Combinatorics},
  volume  = {2025},
  eid     = {8},
  note    = {Article 8},
  year    = {2025},
  doi     = {10.19086/aic.2025.8}
}

@inproceedings{AGL24b,
  author    = {Alrabiah, Omar and Guruswami, Venkatesan and Li, Ray},
  title     = {Randomly punctured {R}eed-{S}olomon codes achieve list-decoding capacity over linear-sized fields},
  booktitle = {Proceedings of the 56th Annual ACM Symposium on Theory of Computing (STOC)},
  pages     = {1458--1469},
  year      = {2024},
  doi       = {10.1145/3618260.3649634}
}

@techreport{BCPZZ26,
  author      = {Brakensiek, Joshua and Chen, Yeyuan and Putterman, Aaron (Louie) and Zhang, Zihan and Zheng, Kai Zhe},
  title       = {Algorithmic list decoding of {R}eed-{S}olomon codes up to capacity},
  institution = {Electronic Colloquium on Computational Complexity (ECCC)},
  number      = {TR26-164},
  year        = {2026},
  url         = {https://eccc.weizmann.ac.il/report/2026/164/},
  note        = {Revision 1, September 2026 (the original version was titled ``\ldots up to capacity in the low-rate regime'')}
}

@book{Ber68,
  author    = {Berlekamp, Elwyn R.},
  title     = {Algebraic Coding Theory},
  publisher = {McGraw-Hill},
  address   = {New York},
  year      = {1968}
}

@inproceedings{BGM23,
  author    = {Brakensiek, Joshua and Gopi, Sivakanth and Makam, Visu},
  title     = {Generic {R}eed-{S}olomon codes achieve list-decoding capacity},
  booktitle = {Proceedings of the 55th Annual ACM Symposium on Theory of Computing (STOC)},
  pages     = {1488--1501},
  year      = {2023},
  doi       = {10.1145/3564246.3585128}
}

@article{BKR10,
  author  = {Ben-Sasson, Eli and Kopparty, Swastik and Radhakrishnan, Jaikumar},
  title   = {Subspace polynomials and limits to list decoding of {R}eed-{S}olomon codes},
  journal = {IEEE Transactions on Information Theory},
  volume  = {56},
  number  = {1},
  pages   = {113--120},
  year    = {2010},
  doi     = {10.1109/TIT.2009.2034780},
  note    = {Preliminary version in FOCS 2006}
}

@misc{CGHLLPS26,
  author       = {Carmon, Dan and Goldberg, Lior and Hab{\"o}ck, Ulrich and Lerer, Leonardo and Lesokhin, Ilya and Papini, Shahar and Samocha, Shahar},
  title        = {{S-two} whitepaper},
  howpublished = {Cryptology ePrint Archive, Paper 2026/532},
  year         = {2026},
  url          = {https://eprint.iacr.org/2026/532}
}

@article{CW07,
  author  = {Cheng, Qi and Wan, Daqing},
  title   = {On the list and bounded distance decodability of {R}eed-{S}olomon codes},
  journal = {SIAM Journal on Computing},
  volume  = {37},
  number  = {1},
  pages   = {195--209},
  year    = {2007},
  doi     = {10.1137/S0097539705447335},
  note    = {Preliminary version in FOCS 2004}
}

@techreport{DKTZ26,
  author      = {Dao, Quang and Kominers, Scott Duke and Thaler, Justin and Zheng, Kai Zhe},
  title       = {Counterexamples to Beyond-{Johnson} Proximity Gaps over Binary Fields},
  institution = {Electronic Colloquium on Computational Complexity (ECCC)},
  number      = {TR26-237},
  year        = {2026},
  url         = {https://eccc.weizmann.ac.il/report/2026/237/}
}

@techreport{Eli57,
  author      = {Elias, Peter},
  title       = {List decoding for noisy channels},
  institution = {Research Laboratory of Electronics, Massachusetts Institute of Technology},
  type        = {Technical Report},
  number      = {335},
  year        = {1957}
}

@article{Eli91,
  author  = {Elias, Peter},
  title   = {Error-correcting codes for list decoding},
  journal = {IEEE Transactions on Information Theory},
  volume  = {37},
  number  = {1},
  pages   = {5--12},
  year    = {1991},
  doi     = {10.1109/18.61123}
}

@article{FKS22,
  author  = {Ferber, Asaf and Kwan, Matthew and Sauermann, Lisa},
  title   = {List-decodability with large radius for {R}eed-{S}olomon codes},
  journal = {IEEE Transactions on Information Theory},
  volume  = {68},
  number  = {6},
  pages   = {3823--3828},
  year    = {2022},
  doi     = {10.1109/TIT.2022.3148779},
  note    = {Preliminary version in FOCS 2021}
}

@inproceedings{GLSTW21,
  author = {Zeyu Guo and Ray Li and Chong Shangguan and Itzhak Tamo and Mary Wootters},
  title = {Improved list-decodability and list-recoverability of {Reed--Solomon} codes via tree packings},
  booktitle = {Proceedings of the 62nd Annual IEEE Symposium on Foundations of Computer Science},
  pages = {708--719},
  year = {2021},
  doi = {10.1109/FOCS52979.2021.00074}
}

@article{GR06,
  author  = {Guruswami, Venkatesan and Rudra, Atri},
  title   = {Limits to list decoding {R}eed-{S}olomon codes},
  journal = {IEEE Transactions on Information Theory},
  volume  = {52},
  number  = {8},
  pages   = {3642--3649},
  year    = {2006},
  doi     = {10.1109/TIT.2006.878164},
  note    = {Preliminary version in STOC 2005}
}

@article{GRS,
  author  = {Oded Goldreich and Ronitt Rubinfeld and Madhu Sudan},
  title   = {Learning Polynomials with Queries: The Highly Noisy Case},
  journal = {SIAM Journal on Discrete Mathematics},
  volume  = {13},
  number  = {4},
  pages   = {535--570},
  year    = {2000},
  doi     = {10.1137/S0895480198344540},
  url     = {https://people.csail.mit.edu/madhu/papers/1995/grs-journ.pdf}
}

@unpublished{GS-johnson,
  author = {Venkatesan Guruswami and Madhu Sudan},
  title  = {Extensions to the {Johnson} Bound},
  note   = {Manuscript},
  month  = feb,
  year   = {2001},
  url    = {https://people.csail.mit.edu/madhu/papers/2001/johnson.pdf}
}

@inproceedings{GS03,
  author = {Venkatesan Guruswami and Igor Shparlinski},
  title = {Unconditional proof of tightness of {Johnson} bound},
  booktitle = {Proceedings of the 14th Annual ACM-SIAM Symposium on Discrete Algorithms},
  pages = {754--755},
  year = {2003}
}

@article{GS99,
  author  = {Guruswami, Venkatesan and Sudan, Madhu},
  title   = {Improved decoding of {R}eed-{S}olomon and algebraic-geometry codes},
  journal = {IEEE Transactions on Information Theory},
  volume  = {45},
  number  = {6},
  pages   = {1757--1767},
  year    = {1999},
  doi     = {10.1109/18.782097},
  note    = {Preliminary version in FOCS 1998}
}

@article{GST23,
  author  = {Goldberg, Eitan and Shangguan, Chong and Tamo, Itzhak},
  title   = {List-decoding and list-recovery of {R}eed-{S}olomon codes beyond the {J}ohnson radius for every rate},
  journal = {IEEE Transactions on Information Theory},
  volume  = {69},
  number  = {4},
  pages   = {2261--2268},
  year    = {2023},
  doi     = {10.1109/TIT.2022.3222877}
}

@inproceedings{Gur02,
  author = {Venkatesan Guruswami},
  title = {Limits to list decodability of linear codes},
  booktitle = {Proceedings of the 34th Annual ACM Symposium on Theory of Computing},
  pages = {802--811},
  year = {2002},
  doi = {10.1145/509907.510022}
}

@article{GV05,
  author = {Venkatesan Guruswami and Alexander Vardy},
  title = {Maximum-likelihood decoding of {Reed--Solomon} codes is {NP}-hard},
  journal = {IEEE Transactions on Information Theory},
  volume = {51},
  number = {7},
  pages = {2249--2256},
  year = {2005},
  doi = {10.1109/TIT.2005.850102}
}

@inproceedings{GZ23,
  author    = {Guo, Zeyu and Zhang, Zihan},
  title     = {Randomly punctured {R}eed-{S}olomon codes achieve the list decoding capacity over polynomial-size alphabets},
  booktitle = {Proceedings of the 64th Annual IEEE Symposium on Foundations of Computer Science (FOCS)},
  pages     = {164--176},
  year      = {2023},
  doi       = {10.1109/FOCS57990.2023.00019}
}

@misc{HKS26,
  author        = {Harsha, Prahladh and Kumar, Mrinal and Saptharishi, Ramprasad},
  title         = {{R}eed-{S}olomon codes at capacity: Algorithmic list-decoding and proximity gaps},
  year          = {2026},
  eprint        = {2610.08610},
  archivePrefix = {arXiv},
  howpublished  = {arXiv:2610.08610},
  url           = {https://arxiv.org/abs/2610.08610}
}

@techreport{Jer26,
  author      = {Jeronimo, Fernando Granha},
  title       = {Algorithmic list decoding at capacity and optimal proximity gaps for {R}eed-{S}olomon codes},
  institution = {Electronic Colloquium on Computational Complexity (ECCC)},
  number      = {TR26-169},
  year        = {2026},
  url         = {https://eccc.weizmann.ac.il/report/2026/169/}
}

@article{JH01,
  author  = {Justesen, J{\o}rn and H{\o}holdt, Tom},
  title   = {Bounds on list decoding of {MDS} codes},
  journal = {IEEE Transactions on Information Theory},
  volume  = {47},
  number  = {4},
  pages   = {1604--1609},
  year    = {2001},
  doi     = {10.1109/18.923744}
}

@article{Joh62,
  author  = {Johnson, Selmer M.},
  title   = {A new upper bound for error-correcting codes},
  journal = {IRE Transactions on Information Theory},
  volume  = {8},
  number  = {3},
  pages   = {203--207},
  year    = {1962},
  doi     = {10.1109/TIT.1962.1057714}
}

@misc{KKH26,
  author       = {Krachun, Dmitry and Kazanin, Stepan and Hab{\"o}ck, Ulrich},
  title        = {Failure of proximity gaps close to capacity},
  howpublished = {Cryptology ePrint Archive, Paper 2026/782},
  year         = {2026},
  url          = {https://eprint.iacr.org/2026/782}
}

@article{Massey69,
  author  = {Massey, James L.},
  title   = {Shift-register synthesis and {BCH} decoding},
  journal = {IEEE Transactions on Information Theory},
  volume  = {15},
  number  = {1},
  pages   = {122--127},
  year    = {1969},
  doi     = {10.1109/TIT.1969.1054260}
}

@article{Peterson60,
  author  = {Peterson, W. Wesley},
  title   = {Encoding and error-correction procedures for the {B}ose-{C}haudhuri codes},
  journal = {IRE Transactions on Information Theory},
  volume  = {6},
  number  = {4},
  pages   = {459--470},
  year    = {1960},
  doi     = {10.1109/TIT.1960.1057586}
}

@article{RS60,
  author  = {Reed, Irving S. and Solomon, Gustave},
  title   = {Polynomial codes over certain finite fields},
  journal = {Journal of the Society for Industrial and Applied Mathematics},
  volume  = {8},
  number  = {2},
  pages   = {300--304},
  year    = {1960},
  doi     = {10.1137/0108018}
}

@inproceedings{RW14,
  author    = {Rudra, Atri and Wootters, Mary},
  title     = {Every list-decodable code for high noise has abundant near-optimal rate puncturings},
  booktitle = {Proceedings of the 46th Annual ACM Symposium on Theory of Computing (STOC)},
  pages     = {764--773},
  year      = {2014},
  doi       = {10.1145/2591796.2591797}
}

@article{Sch20,
  author = {Kai-Uwe Schmidt},
  title = {Quadratic and symmetric bilinear forms over finite fields and their association schemes},
  journal = {Algebraic Combinatorics},
  volume = {3},
  number = {1},
  pages = {161--189},
  year = {2020}
}

@article{Sin64,
  author = {Richard C. Singleton},
  title = {Maximum distance {$q$}-nary codes},
  journal = {IEEE Transactions on Information Theory},
  volume = {10},
  number = {2},
  pages = {116--118},
  year = {1964},
  doi = {10.1109/TIT.1964.1053661}
}

@article{ST20,
  author  = {Shangguan, Chong and Tamo, Itzhak},
  title   = {Generalized {S}ingleton bound and list-decoding {R}eed-{S}olomon codes beyond the {J}ohnson radius},
  journal = {SIAM Journal on Computing},
  volume  = {52},
  number  = {3},
  pages   = {684--717},
  year    = {2023},
  doi     = {10.1137/20M138795X},
  note    = {Preliminary version, titled ``Combinatorial list-decoding of Reed-Solomon codes beyond the Johnson radius,'' in STOC 2020}
}

@article{Sud97,
  author  = {Sudan, Madhu},
  title   = {Decoding of {R}eed {S}olomon codes beyond the error-correction bound},
  journal = {Journal of Complexity},
  volume  = {13},
  number  = {1},
  pages   = {180--193},
  year    = {1997},
  doi     = {10.1006/jcom.1997.0439}
}

@misc{WB86,
  author       = {Welch, Lloyd R. and Berlekamp, Elwyn R.},
  title        = {Error correction for algebraic block codes},
  howpublished = {U.S. Patent 4,633,470},
  month        = dec,
  year         = {1986}
}

@article{Woz58,
  author  = {Wozencraft, John M.},
  title   = {List decoding},
  journal = {Quarterly Progress Report, Research Laboratory of Electronics, Massachusetts Institute of Technology},
  volume  = {48},
  pages   = {90--95},
  year    = {1958}
}

@misc{Zhu26,
  author        = {Zhu, Minshen},
  title         = {Superpolynomial Lists for {Reed--Solomon} Codes over Binary Extension Fields},
  year          = {2026},
  eprint        = {2610.15815},
  archivePrefix = {arXiv},
  howpublished  = {arXiv:2610.15815},
  url           = {https://arxiv.org/abs/2610.15815}
}

@article{ZP81,
  author  = {Zyablov, Victor V. and Pinsker, Mark S.},
  title   = {List concatenated decoding},
  journal = {Problems of Information Transmission},
  volume  = {17},
  number  = {4},
  pages   = {236--240},
  year    = {1981}
}

\end{document}